\documentclass[amsmath,amssymb,reprint,superscriptaddress]{revtex4-1}

\usepackage[utf8]{inputenc}
\usepackage[T1]{fontenc}
\usepackage{mathptmx}

\usepackage{graphicx}
\usepackage[italicdiff]{physics}
\usepackage{siunitx}
\usepackage{mhchem}

\usepackage{xparse}
\NewDocumentCommand\drv{m m}{f_{k_{#1}k_{#2}}}
\NewDocumentCommand\ddrv{m m}{\kappa_{k_{#1}k_{#2}}}

\NewDocumentCommand\rij{}{r_{ij}}
\NewDocumentCommand\sij{}{\sigma_{ij}}

\NewDocumentCommand\Melem{m o m o}{
{
 \IfValueTF{#2}
  {
   \IfValueTF{#4}
    {
     \mathcal{M}_{{#1}_{#3} {#2}_{#4}}
    }
    {
     \mathcal{M}_{{#1}_{#3} {#2}_{#3}}
    }
  }
  {
   \IfValueTF{#4}
    {
     \mathcal{M}_{{#1}_{#3} {#1}_{#4}}
    }
    {
     \mathcal{M}_{{#1}_{#3} {#1}_{#3}}
    }
  }
}
}
\NewDocumentCommand\rkikj{m m}{r_{k_{#1} k_{#2}}}
\NewDocumentCommand\xkikj{m m}{X_{k_{#1} k_{#2}}}
\NewDocumentCommand\ykikj{m m}{Y_{k_{#1} k_{#2}}}
\NewDocumentCommand\zkikj{m m}{Z_{k_{#1} k_{#2}}}
\NewDocumentCommand\kf{m m}{\pqty{\ddrv{#1}{#2} + \frac{\drv{#1}{#2}}{\rkikj{#1}{#2}}}}
\NewDocumentCommand\drdtheta{m m}{\frac{\pqty{\alpha-1}bn_{#1}}{2\rkikj{#1}{#2}\sqrt{I_1}} \pqty{\xkikj{#1}{#2}\cos\theta_{#1}\sin\phi_{#1} - \ykikj{#1}{#2}\cos\theta_{#1}\cos\phi_{#1} - \zkikj{#1}{#2}\sin\theta_{#1}}}
\NewDocumentCommand\drdthetaj{m m}{\frac{\pqty{\alpha-1}bn_{#2}}{2\rkikj{#1}{#2}\sqrt{I_1}} \pqty{-\xkikj{#1}{#2}\cos\theta_{#2}\sin\phi_{#2} + \ykikj{#1}{#2}\cos\theta_{#2}\cos\phi_{#2} + \zkikj{#1}{#2}\sin\theta_{#2}}}
\NewDocumentCommand\drdphi{m m}{\frac{\pqty{\alpha-1}bn_{#1}}{2\rkikj{#1}{#2}\sqrt{I_1}\sin\theta_{#1}} \pqty{\xkikj{#1}{#2}\sin\theta_{#1}\cos\phi_{#1} + \ykikj{#1}{#2}\sin\theta_{#1}\sin\phi_{#1}}}
\NewDocumentCommand\drdphij{m m}{\frac{\pqty{\alpha-1}bn_{#2}}{2\rkikj{#1}{#2}\sqrt{I_1}\sin\theta_{#2}} \pqty{-\xkikj{#1}{#2}\sin\theta_{#2}\cos\phi_{#2} - \ykikj{#1}{#2}\sin\theta_{#2}\sin\phi_{#2}}}

\usepackage{xcolor,hyperref}
 \hypersetup{
 colorlinks,
 linkcolor={blue!50!black},
 citecolor={blue!50!black},
 urlcolor={blue!80!black}
}

\begin{document}
\title{Mechanical and Vibrational Properties of Three-Dimensional Dimer Packings Near the Jamming Transition}
\author{Kumpei Shiraishi}
\email{kumpeishiraishi@g.ecc.u-tokyo.ac.jp}
\affiliation{Graduate School of Arts and Sciences, University of Tokyo, Komaba, Tokyo 153-8902, Japan}
\author{Hideyuki Mizuno}
\affiliation{Graduate School of Arts and Sciences, University of Tokyo, Komaba, Tokyo 153-8902, Japan}
\author{Atsushi Ikeda}
\affiliation{Graduate School of Arts and Sciences, University of Tokyo, Komaba, Tokyo 153-8902, Japan}
\affiliation{Research Center for Complex Systems Biology, Universal Biology Institute, University of Tokyo, Komaba, Tokyo 153-8902, Japan}
\date{\today}

\begin{abstract}
We comprehensively study mechanical and vibrational properties of dimer packings in three-dimensional space with particular attention on critical scaling behaviors near the jamming transition.
First, we confirm the dependence of the packing fraction at the transition on the aspect ratio, the isostatic contact number at the transition, and the scaling dependence of the excess contact number on the excess density.
Second, we study the elastic moduli, bulk and shear moduli, and establish power-law scaling of them.
Finally, we study the vibrational density of states and its characteristic frequency.
The vibrational density of states shows two plateaus in the lower- and higher-frequency regions, which are characterized by rotational and translational vibrational modes, respectively.
The onset frequency of the lower-frequency plateau scales linearly to the square root of excess density.
The scaling laws in the mechanical and vibrational properties are consistent between two- and three-dimensional dimers, and they are identical to those in spheres.
\end{abstract}

\maketitle

\section{Introduction}
Glasses exhibit low-temperature thermal properties distinct from those of crystals, and understanding the behavior has been a long-standing problem in the field of condensed matter physics.
At low temperature, the specific heat $C$ of crystals follows the universal law $C \propto T^3$, where $T$ is the temperature, which is well understood by the Debye theory.
In contrast, the specific heat of glasses follows $C \propto T$ at $T \sim \SI{1}{\kelvin}$.\cite{Zeller_Pohl_1971}
When the temperature increases, the specific heat of glasses approaches the behavior of the Debye theory, but $C/T^3$ displays a peak at $T \sim \SI{10}{\kelvin}$, which is called the boson peak.
These anomalous thermal properties stem from the anomalous vibrational properties of glasses.
The vibrational density of states (VDOS) $g\pqty{\omega}$ of glasses reduced by the Debye prediction $\omega^2$ shows the boson peak at $\omega \sim \SI{1}{\tera\hertz}$, i.e., glasses have excess vibrational modes in this frequency region.\cite{Buchenau_1984}
The boson peak phenomenon has been universally observed in amorphous solids regardless of their constituents.

Numerous theoretical and numerical studies were conducted to understand the anomalous vibrational properties of glasses.
Those studies often modeled the constituents of glasses based on spherical particles, i.e., particles that interact via spherically symmetric potentials.
In particular, the understanding of the vibrational properties of glasses has progressed through studies on athermal soft repulsive spheres.\cite{O_Hern_2003,van_Hecke_2010}
The system shows a liquid-like property with low packing fraction $\varphi$.
When the system is compressed, it reaches a packing fraction $\varphi_J$, where it solidifies through the jamming transition.
The number of contacts per particle at the transition takes an isostatic value $z_\text{iso} = 2d$, where $d$ is the spatial dimension.\cite{O_Hern_2003}
The excess contact number $\Delta z = z - z_\text{iso}$ shows the critical scaling with excess density $\Delta\varphi = \varphi-\varphi_J$: $\Delta z \propto \Delta\varphi^{1/2}$.\cite{O_Hern_2003}
The particles of the packing are in an amorphous structure; therefore, non-affine relaxation occurs in response to the affine deformation.
The shear modulus is affected by the non-affine relaxation, which leads to the critical scaling $G \propto \Delta z$.\cite{O_Hern_2003,Wyart_2005,Zaccone_2011,Cui_2019}
The VDOS of the system shows a plateau in the low-frequency region, i.e., excess vibrational modes compared to the Debye prediction.\cite{O_Hern_2003}
The onset frequency of the plateau, which is denoted as $\omega^*$, is proportional to the square root of the excess density,\cite{Silbert_2005} and the excess contact number.\cite{Wyart_PRE_2005}
This behavior is theoretically explained by the variational argument~\cite{Yan_2016} and effective medium theory.\cite{Wyart_EPL_2010}

In contrast to theoretical and numerical studies, the glasses in experimental studies are normally composed of complex-shaped, nonspherical particles.
For example, molecular glasses are generally composed of molecules with a large number of atoms because simple-shaped molecules, e.g., diatomic or triatomic molecules, easily crystallize.
To generate glasses with simple-shaped molecules, rapid cooling is required to prevent crystallization.
The vapor-deposition method, which achieves rapid cooling and highly stable glasses,\cite{Ediger_2017} can generate glasses of simple molecules such as \ce{H2O},\cite{Narten_1976} \ce{CCl4},\cite{Wenzel_1976} \ce{CS2},\cite{Yamamuro_2003} and \ce{CO2},\cite{Yamamuro_2016} but has never succeeded in generating glasses of spherical monatomic elements.
Colloidal particles can formed as a dumbbell shape~\cite{Kim_2006} and other various nonspherical shapes.\cite{Sacanna_2011}
Protein cores are jammed structure of amino acids, where the packing fraction is approximately \num{0.56}.\cite{Gaines_2016}
To reproduce this packing fraction, amino acids must be modeled as bumpy, non-axisymmetric, nonspherical particles, instead of smooth nonspherical particles such as ellipsoids and spherocylinders.\cite{Gaines_2017}
Therefore, to understand the experimental results, it is necessary to extend the theoretical and numerical studies to the nonspherical cases.

In this work, we consider the random packings of three-dimensional (3D) dimer particles as a prototypical example of glasses composed of anisotropic particles.
The dimers are composed of two spheres with a rigid bond.
In particular, we use numerical simulations to study the critical behaviors of the geometrical, mechanical, and vibrational properties of this model near the jamming transition.
Theoretically, Baule et al.~developed a geometric mean-field theory of random packings of axisymmetric particles, which gives a prediction of the packing density and contact numbers of 3D dimers.\cite{Baule_2013}
However, there is no theory for the mechanical and vibrational properties of 3D dimers.
In numerical simulations, the jamming transition of two-dimensional (2D) dimers has been thoroughly studied.\cite{Schreck_2010,Shiraishi_2019}
These studies reveal that the contact number becomes the isostatic value at the jamming transition, but a certain type of rattler particles, called rotational rattlers, must be taken into account to define the isostatic contact number.\cite{Shiraishi_2019}
The shear modulus of 2D dimers follows the critical behavior of spherical particles.\cite{Schreck_2010}
The VDOS of 2D dimers has a peak and two plateaus, and rotational motions are dominant at the peak and in the lower-frequency plateau.\cite{Shiraishi_2019}
The onset frequency of the lower-frequency plateau obeys the same critical law as spherical particles.\cite{Shiraishi_2019}

However, it is not clear whether these characteristic behaviors of 2D dimers remain valid in 3D dimers, which is a more realistic system than 2D dimers.
It is worth to note that studying dimers, particularly 3D dimers, is technically demanding.
For dimers~(or particles with anisotropic shapes), it is necessary to solve the equations of motions, which are composed of intricately coupled translational and rotational degrees of freedoms (DOFs), and implement a cumbersome vibrational analysis.
We performed this complex analysis on 3D dimers and present a comprehensive report on the mechanical and vibrational properties of the packings.

\section{Methods}\label{sec:methods}
\subsection{Equations of motions and numerical protocols}
Our model of a 3D dimer is a composite of two spheres with identical diameters, connected by a bond with constant length.
We refer to a sphere that composes a dimer as a monomer.
Consider a dimer $i$ in 3D space whose translational and rotational DOFs are designated by $\vb{r}_i = \pqty{x_i, y_i, z_i, \phi_i, \theta_i, \psi_i}$, where $x_i$, $y_i$ and $z_i$ are the positions of the center of gravity of the dimer, and $\phi_i$, $\theta_i$ and $\psi_i$ are the Euler angles, which are defined by the $z$-$x$-$z$ convention.\cite{Landau_Mechanics}
Therefore, the range of the angles are $0 \leq \theta_i < \pi$ and $0 \leq \phi_i,\psi_i < 2\pi$.
The length of the major axis is set to $a_i$, and the length of the minor axis is set to $b_i$.
The shape of the dimer is described using the aspect ratio $\alpha \equiv a_i/b_i$.
Let us consider a packing composed of $N$ dimers.
We consider a monodisperse system; therefore, all aspect ratios and lengths of the minor axis are identical to $\alpha$ and $\sigma$, respectively, throughout the system.
All particles have equal mass $m$.

We first derive the translational and rotational equations of motion of the system.
The potential energy of the system is given by
\begin{align}
 V = \sum_{i=1}^N\sum_{j=i+1}^N \sum_{n_i = \pm 1}\sum_{n_j = \pm 1} \frac{\epsilon}{2}\pqty{1 - \frac{\rkikj{i}{j}}{\sij}}^2 H\pqty{1 - \frac{\rkikj{i}{j}}{\sij}}, \label{potential}
\end{align}
where
\begin{align}
\begin{aligned}
 \rkikj{i}{j} &= \sqrt{\xkikj{i}{j}^2 + \ykikj{i}{j}^2 + \zkikj{i}{j}^2}, \\
 \xkikj{i}{j} &= x_i - x_j + \frac{b\pqty{\alpha-1}}{2}\pqty{ n_i\sin\theta_i\sin\phi_i - n_j\sin\theta_j\sin\phi_j}, \\
 \ykikj{i}{j} &= y_i - y_j + \frac{b\pqty{\alpha-1}}{2}\pqty{-n_i\sin\theta_i\cos\phi_i + n_j\sin\theta_j\cos\phi_j}, \\
 \zkikj{i}{j} &= z_i - z_j + \frac{b\pqty{\alpha-1}}{2}\pqty{ n_i\cos\theta_i           - n_j\cos\theta_j}, \label{rkikj}
\end{aligned}
\end{align}
$H\pqty{x}$ is the Heaviside step function, i.e., $H\pqty{x} = 1$ for $x \geq 0$ and $H\pqty{x} = 0$ for $x < 0$, $\epsilon$ is the characteristic energy scale, $b$ is the length of the minor axis (monomer diameter), and $k_i \equiv \pqty{i,n_i}$ is a set of integers, which designates a monomer of dimer $i$.
We define directions 1, 2, and 3 as the principal axes of inertia and $I_1$, $I_2$, and $I_3$ as the principal moments of inertia.
See Supplemental Material for the calculation.
The angular velocity in these directions is described by the time derivatives of the Euler angles as
\begin{align}
 \Omega_i = \pmqty{\omega_{i,1} \\ \omega_{i,2} \\ \omega_{i,3}} 
          = \pmqty{\dot{\theta}_i\cos\psi_i + \dot{\phi}_i\sin\psi_i\sin\theta_i \\ -\dot{\theta}_i\sin\psi_i + \dot{\phi}_i\cos\psi_i\sin\theta_i \\ \dot{\psi}_i + \dot{\phi}_i\cos\theta_i}.
\end{align}
Using this expression, the kinetic energy of the system is written as
\begin{align}
 T = \sum_{i=1}^N\bqty{\frac{m}{2}\pqty{\dot{x}_i^2 + \dot{y}_i^2 + \dot{z}_i^2} + \frac{1}{2}\pqty{I_1\omega_{i,1}^2 + I_2\omega_{i,2}^2 + I_3\omega_{i,3}^2}},
\end{align}
so the Lagrangian $\mathcal{L}$ of the system is $\mathcal{L} = T - V$.

Now we consider the equations of motion of the system.
The translational parts of Euler-Lagrange equations are
\begin{align}
 \dv{t}\pdv{\mathcal{L}}{\dot{x}_i} = \pdv{\mathcal{L}}{x_i}, \quad
 \dv{t}\pdv{\mathcal{L}}{\dot{y}_i} = \pdv{\mathcal{L}}{y_i}, \quad
 \dv{t}\pdv{\mathcal{L}}{\dot{z}_i} = \pdv{\mathcal{L}}{z_i}
\end{align}
which can be reduced to
\begin{align}
 m\ddot{x}_i = -\pdv{V}{x_i}, \quad
 m\ddot{y}_i = -\pdv{V}{y_i}, \quad
 m\ddot{z}_i = -\pdv{V}{z_i}. \label{eom_tr}
\end{align}
The rotational parts are
\begin{align}
 \dv{t}\pdv{\mathcal{L}}{\dot{\phi}_i}   = \pdv{\mathcal{L}}{\phi_i}, \quad
 \dv{t}\pdv{\mathcal{L}}{\dot{\theta}_i} = \pdv{\mathcal{L}}{\theta_i}, \quad
 \dv{t}\pdv{\mathcal{L}}{\dot{\psi}_i}   = \pdv{\mathcal{L}}{\psi_i}. \label{Lagrange_eq_rot}
\end{align}
Since two of the principal moments of inertia are equal ($I_1 = I_2 \neq I_3$, see Supplemental Material for the proof), 3D dimer is a symmetrical top.
In addition, Eqs.~\eqref{potential} and \eqref{rkikj} show that the potential does not depend on $\psi$: $\dd{V}/\dd{\psi} \equiv 0$.
Using these facts and rearranging Eqs.~\eqref{Lagrange_eq_rot}, we obtain Euler's equations of the system:
\begin{align}
\begin{aligned}
 I_1\dot{\omega}_{i,1} &= N_{i,1} + \pqty{I_1 - I_3}\omega_{i,2}\omega_{i,3}, \\
 I_1\dot{\omega}_{i,2} &= N_{i,2} + \pqty{I_3 - I_1}\omega_{i,3}\omega_{i,1}, \\
 I_3\dot{\omega}_{i,3} &= N_{i,3}, \label{eom_rot}
\end{aligned}
\end{align}
where
\begin{align}
\begin{aligned}
 N_{i,1} &= -\frac{\sin\psi_i}{\sin\theta_i}\pdv{V}{\phi_i} - \cos\psi_i\pdv{V}{\theta_i}, \\
 N_{i,2} &= -\frac{\cos\psi_i}{\sin\theta_i}\pdv{V}{\phi_i} + \sin\psi_i\pdv{V}{\theta_i}, \\
 N_{i,3} &= 0
\end{aligned}
\end{align}
are the components of the torque acting on dimer $i$.

To numerically solve the equations of motion~\eqref{eom_tr} and~\eqref{eom_rot}, we implemented the fifth-order Gear predictor-corrector method~\cite{Allen_Tildesley} for the translational DOFs and the fourth-order procedure for the rotational DOFs.
See Appendix~A in the literature on liquid water~\cite{Rahman_Stillinger_1971} for the implementation details.
In our implementation, the quaternion parameters were used instead of the Euler angles when following the time evolution of the rotational DOFs.\cite{Evans_Murad_1977}

We first prepared a random configuration of $N=\num{500}$ at sufficiently small packing fraction $\varphi_0 = \num{0.2}$.
Except for the data at the jamming transition in Sects.~\ref{sec:phiJ} and \ref{sec:contact}, we performed the calculation for over \num{100} samples.
The packing fraction for monodisperse $N$ dimers in the simulation box with system length $L$ is given as
\begin{align}
 \varphi = \frac{N\pi\alpha^2\pqty{3-\alpha}b^3}{12L^3}.
\end{align}
See Supplemental Material for the calculation of the volume of a 3D dimer.
We employed the periodic boundary condition on the system.
We minimized the potential energy of the random configuration using the FIRE algorithm\cite{Bitzek_2006} and gradually compressed the system in increments of $\delta\varphi=\num{e-3}$.
The jamming transition point, which is defined as the point at which the potential energy per particle is in the range of $\num{1e-16} < V/N < \num{2e-16}$, was explored by the bisection method.\cite{Gao_2006}
Once the configuration at the transition was obtained, we compressed the system and obtained the configuration at each excess density $\Delta\varphi = \varphi - \varphi_J$.
The energy minimization was terminated when either of the following two conditions was satisfied:\cite{Gao_2006}
(i) the total potential per particle is sufficiently small $V/N < \num{e-16}$ or 
(ii) the total potential energies of two successive minimization steps $t$ and $t+1$ are nearly identical: $\abs{V_{t+1}-V_t}/V_t < \num{e-16}$.
We changed the termination condition when the system was above the transition.
The maximum absolute values of the force and the torque acting on each particle were calculated, and if the summation of the two was below \num{e-12}, then the minimization was terminated.
In the obtained packings, we recursively removed rattler particles with fewer contacts than the number of DOFs, $d_f \equiv 5$.
In the following, the number of particles without rattler particles is denoted as $N$.
We use $m$, $\sigma$, and $\epsilon$ as units of mass, length, and energy, respectively.
The frequency is measured by $\pqty{m\sigma^2/\epsilon}^{-1/2}$.
The pressure, stress, and modulus, which will be defined in Sect.~\ref{sec:mechprop}, are measured in units of $\epsilon/\sigma^d$, where $d \equiv 3$ is the spatial dimension.

\subsection{Vibrational analysis}
Vibrational analysis was performed on the dimer packings obtained by the above protocol.
To formulate the analysis, we consider the linearization of the equations of motion.
First, we linearize the left-hand sides of Eq.~\eqref{Lagrange_eq_rot} around the equilibrium orientations:
\begin{align}
\begin{aligned}
 \dv{t}\pdv{\mathcal{L}}{\dot{\phi}_i}   &\simeq \pqty{I_1\sin^2\theta_i + I_3\cos^2\theta_i}\ddot{\phi}_i + I_3\cos\theta_i\ddot{\psi}_i, \\
 \dv{t}\pdv{\mathcal{L}}{\dot{\theta}_i} &=      I_1\ddot{\theta}_i, \\
 \dv{t}\pdv{\mathcal{L}}{\dot{\psi}_i}   &\simeq I_3\cos\theta_i\ddot{\phi}_i + I_3\ddot{\psi}_i. \label{linearized_eom_psi}
\end{aligned}
\end{align}
Since the potential~\eqref{potential} does not depend on $\psi$, the right-hand side of the $\psi$-component of Eq.~\eqref{Lagrange_eq_rot} is zero.
Therefore, by solving the equation in which the $\psi$-component of Eq.~\eqref{linearized_eom_psi} is equal to zero, we obtain
\begin{align}
 \ddot{\psi}_i = -\ddot{\phi}_i\cos\theta_i.
\end{align}
This relation leads to the linearized equations of motion, that is
\begin{align}
\begin{aligned}
 \ddot{x}_i = -\pdv{V}{x_i}, \quad
 \ddot{y}_i = -\pdv{V}{y_i}, \quad
 \ddot{z}_i = -\pdv{V}{z_i}
\end{aligned}
\end{align}
for the translational DOFs (we set $m$ as unity) and
\begin{align}
\begin{aligned}
 I_1\sin^2\theta_i\ddot{\phi_i} = -\pdv{V}{\phi_i}, \quad
 I_1\ddot{\theta_i}		= -\pdv{V}{\theta_i}
\end{aligned}
\end{align}
for the rotational DOFs.
Therefore, we employ the normal coordinate of dimer $i$ of $\tilde{\vb{r}}_i = \pmqty{x_i & y_i & z_i & \tilde{\phi}_i & \tilde{\theta}_i}$, where $\tilde{\phi}_i = \sqrt{I_1}\sin\theta_i\phi_i$ and $\tilde{\theta}_i = \sqrt{I_1}\theta_i$, and that of the entire system of $\tilde{\vb{r}} = \pmqty{\tilde{\vb{r}}_1 & \cdots & \tilde{\vb{r}}_N}$.
Suppose harmonic vibrations around the equilibrium coordinate $\tilde{\vb{r}}_0$; then, the equation of motion is described as
\begin{align}
 \dv[2]{\tilde{\vb{u}}}{t} = -\mathcal{M}\tilde{\vb{u}},
\end{align}
where $\tilde{\vb{u}} = \tilde{\vb{r}} - \tilde{\vb{r}}_0$ is the displacement from $\tilde{\vb{r}}_0$.
Matrix $\mathcal{M}$ is called the dynamical matrix, and its elements are
\begin{align}
 \Melem{\tilde{\vb{r}}}{i}[j] = \pdv{V}{\tilde{\vb{r}}_i}{\tilde{\vb{r}}_j}. \label{dynmat_def}
\end{align}
The explicit elements of this matrix are given in Supplemental Material.

By calculating the matrix for each configuration and diagonalizing it by LAPACK,\cite{lapack} we obtained a set of eigenvalues $\lambda_k$ and corresponding eigenvectors $\vb{e}_k \equiv \pmqty{\vb{e}_k^1 & \cdots & \vb{e}_k^N}$ (size $d_fN$), where $\vb{e}^i_k \equiv \pmqty{e^{i,x}_k & e^{i,y}_k & e^{i,z}_k & e^{i,\tilde{\phi}}_k & e^{i,\tilde{\theta}}_k}$, and $k = 1,\dots,d_fN$.
The eigenfrequencies are given as $\omega_k = \sqrt{\lambda_k}$, and the eigenvectors are orthonormalized as $\vb{e}_k \vdot \vb{e}_l \equiv \sum_i \vb{e}_k^i \vdot \vb{e}_l^i = \delta_{kl}$ where $\delta_{kl}$ is the Kronecker delta.

\subsection{Unstressed system}\label{sec:method_unstress}
In this work, we performed vibrational eigenmode analysis of the ``unstressed system'' in addition to the original system.
Since the monomers are modeled by harmonic spheres in the present model, the forces between the monomers are always repulsive.
Thus, we refer to this original state as the stressed system.
In the unstressed system, we retain all stiffnesses~(i.e., the second derivative of the potential) between monomers but drop the force in the analysis.
In other words, the unstressed system is the model which all contacts in the original system are replaced with relaxed springs.
For the case of sphere packings, the theoretical understanding was first constructed based on the unstressed system,\cite{Wyart_EPL_2005,Wyart_EPL_2010} which was subsequently extended to the stressed system by considering the effects of the forces.\cite{Wyart_PRE_2005,DeGiuli_2014}
The unstressed system of sphere packings exhibits a characteristic plateau in the VDOS and critical behavior of this plateau near the jamming transition.
The forces in sphere packings do not affect this plateau and its critical behavior, whereas they make the system mechanically unstable~\cite{Lerner_2014} and alter the nature of the very low-frequency vibrational modes.\cite{Mizuno_2017}
In particular, quasilocalized vibrational modes are induced by the repulsive forces.\cite{Mizuno_2017,Lerner_2018,Shimada_Mizuno_Wyart_Ikeda_2018}
In this work, we study the role of repulsive forces in the dimer packings.

\section{Results}
\subsection{Jamming density}\label{sec:phiJ}
\begin{figure}[tb]
 \centering
 \includegraphics[width=\linewidth]{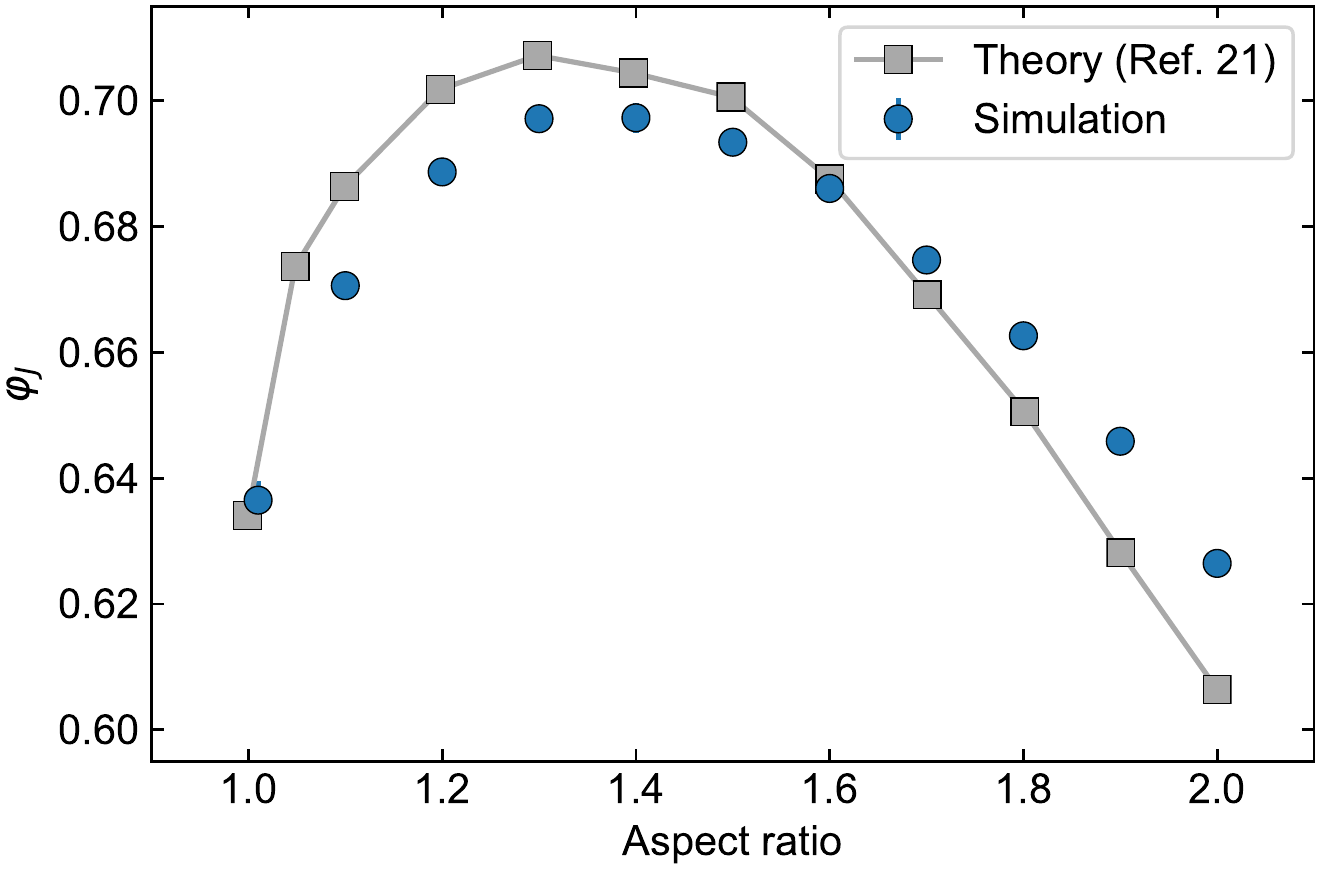}
 \caption{Packing fraction at the jamming transition of 3D dimers. The theoretical prediction is extracted from Ref.~\onlinecite{Baule_2013}. The simulation data are obtained by averaging over at least 10 samples for each aspect ratio.}
 \label{fig:phiJ}
\end{figure}

We generated 3D dimer packings at the jamming transition for various aspect ratios.
The packing fraction at the jamming $\varphi_J$ as a function of $\alpha$ is shown in Fig.~\ref{fig:phiJ}.
When the aspect ratio is close to unity, the density is close to the packing density of random close packings of spheres $\varphi_J \simeq \SI{64}{\percent}$.
The density increases when the aspect ratio increases, and takes the maximum value at approximately $\alpha \simeq \num{1.3}$.
After $\alpha$ exceeds \num{1.4}, $\varphi_J$ decreases with increasing $\alpha$.
Interestingly, the jamming density of 3D dimers with $\alpha=\num{2.0}$ is lower than that of the random close packings of spheres, although the packings are seemingly composed of spheres.
We also plotted the jamming density predicted by the geometric mean-field theory developed by Baule et al.\cite{Baule_2013}
Clearly, our numerical results of the jamming density are comparable to the theoretical predictions.
In particular, there is a quantitative consistency in the value and location of the maximal packing density.

\subsection{Contact number}\label{sec:contact}
\begin{figure}[tb]
 \centering
 \includegraphics[width=\linewidth]{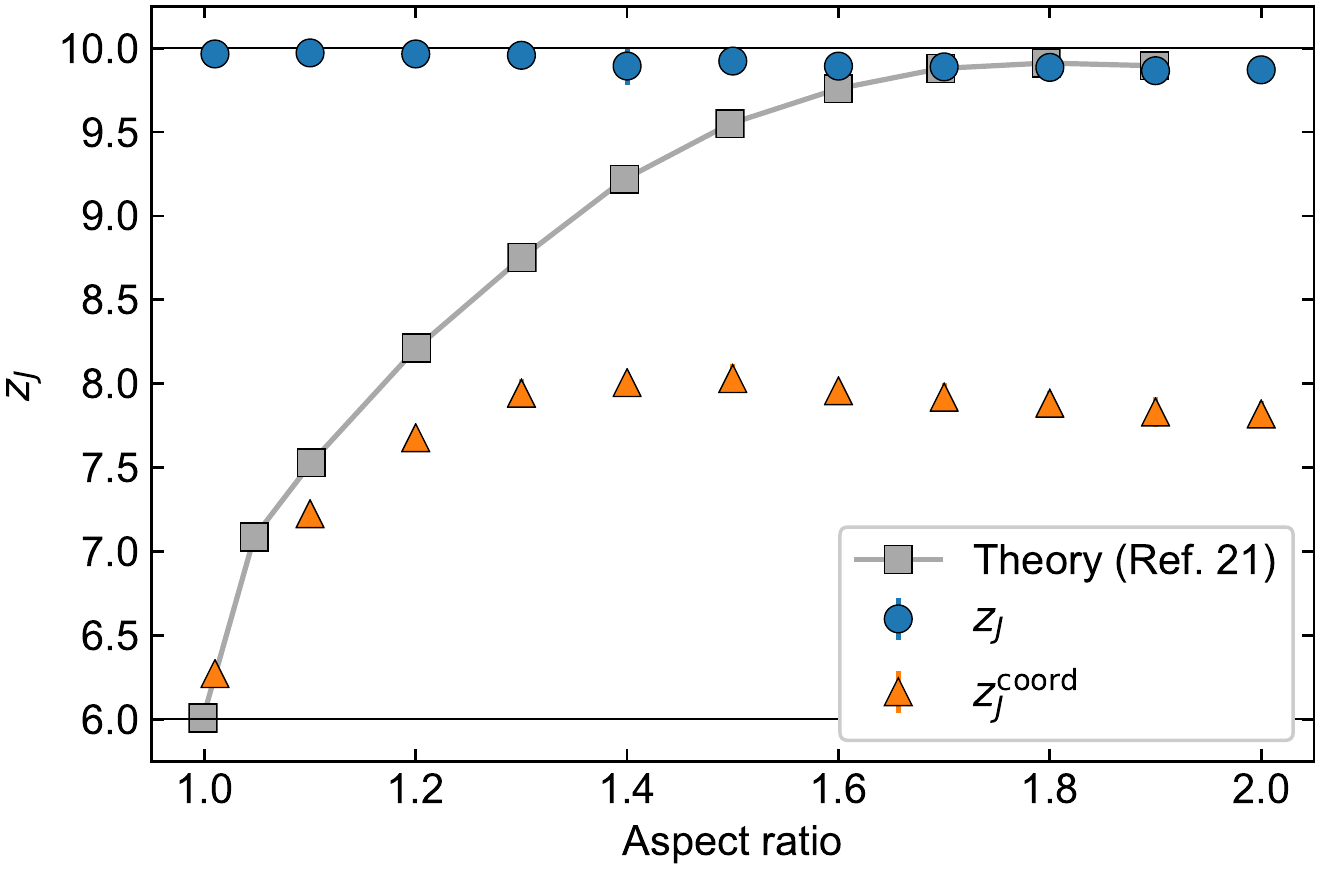}
 \caption{Number of contacts per particle $z_J$ and number of neighboring particles $z_J^\text{coord}$ in the configurations at the jamming. The theoretical prediction of $z_J^\text{coord}$ is extracted from Ref.~\onlinecite{Baule_2013}. The simulation data are obtained by averaging over at least 10 samples for each aspect ratio.}
 \label{fig:zJ}
\end{figure}

\begin{figure}[tb]
 \centering
 \includegraphics[width=\linewidth]{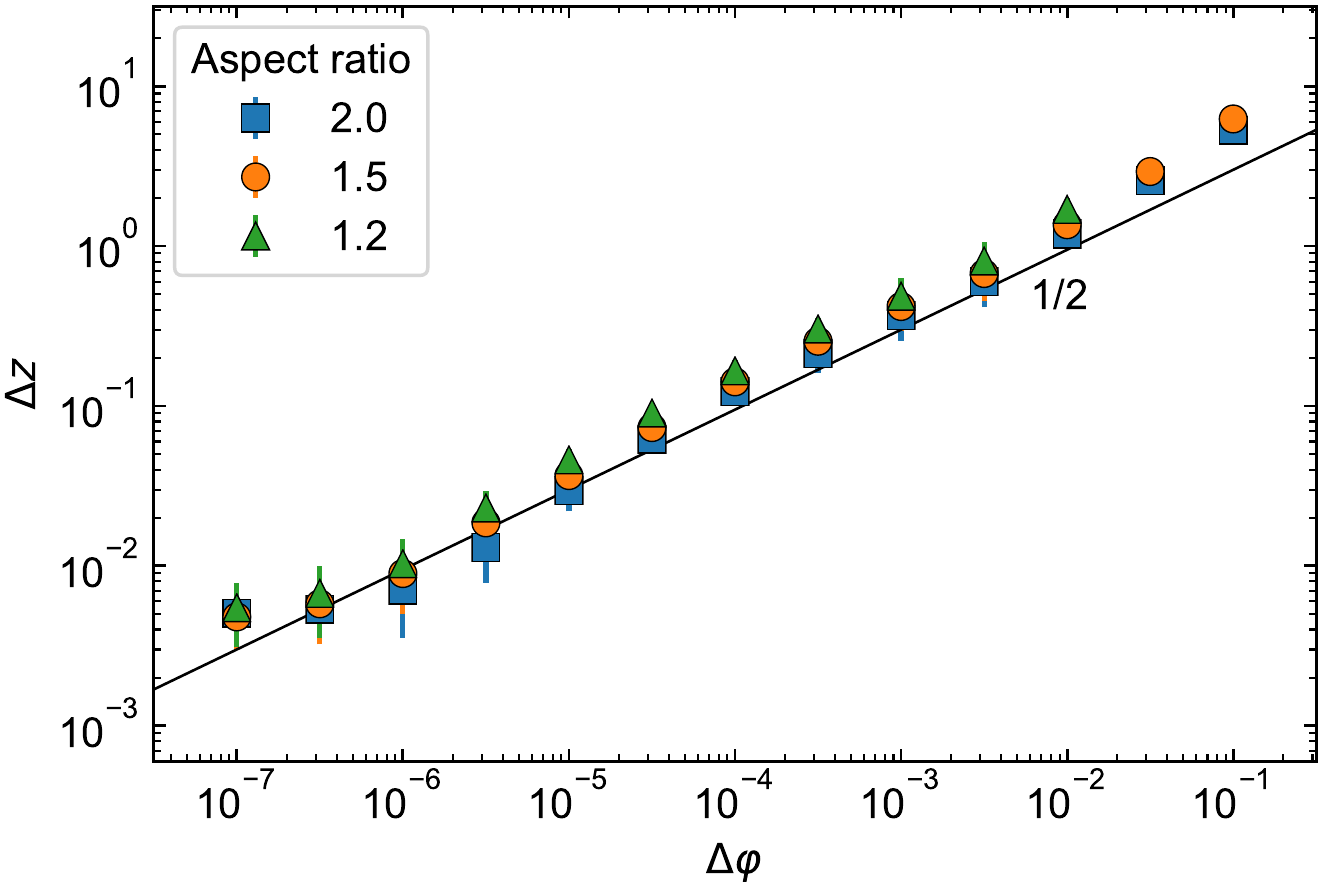}
 \caption{Excess contact number $\Delta z$ as a function of excess density $\Delta\varphi$ for 3D dimers with $\alpha=\num{1.2},\num{1.5},\num{2.0}$. $\Delta z$ shows the same critical behavior $\Delta z \propto \Delta\varphi^{1/2}$ for each aspect ratio.}
 \label{fig:dz}
\end{figure}

Next, we analyzed the contact number and coordination number of 3D dimer packings at the jamming transition (Fig.~\ref{fig:zJ}).
First, we calculated the contact number $z_J$, which is defined as the number of all contacts in the packing divided by the number of dimer particles.
For each aspect ratio, the values of $z_J$ are approximately 10, which is twice the number of DOFs $d_f \equiv 5$, suggesting that the dimer packings are isostatic.
This point will be studied later.
We also calculated the coordination number $z_J^\text{coord}$, which is defined as the number of contacting pairs of dimers divided by the number of dimers.
The contact number and coordination number are equivalent for convex particles such as spheres, spherocylinders, and ellipsoids but not for concave particles such as dimers since one particle can have multiple contacts with a neighboring particle in the latter case.
Figure~\ref{fig:zJ} shows that $z_J^\text{coord}$ approaches 6 as $\alpha \to \num{1}$, which is the isostatic number of spherical particles, i.e., twice the spatial dimension.
When $\alpha$ deviates from unity, $z_J^\text{coord}$ continuously increases and converges to 8.
Therefore, all of these results imply that one dimer has eight neighboring dimers, one contact with each of six neighboring dimers and two contacts with each of two neighboring dimers on average.
Interestingly, $z_J^\text{coord} \approx \num{8}$ at large aspect ratios has been observed for random packings of 3D spherocylinders,\cite{Wouterse_2009,Zhao_2012} where the particles have many contacts on their cylindrical part.
Our results suggest that there is a geometrical similarity between the random packings of 3D dimers and spherocylinders.
The theory of Baule et al.~\cite{Baule_2013} predicts saturation of the coordination number for 3D spherocylinders, but not for 3D dimers, as indicated by the solid line in Fig.~\ref{fig:zJ}.

Finally, we quantitatively analyzed the critical behaviors of the contact numbers.
To this end, we introduce the precise definition of isostatic contact number $z^N_\text{iso}$ for 3D dimers.
According to the discussion on 2D dimers,\cite{Shiraishi_2019} there is a certain type of rattler particle called a rotational rattler, which possesses all of the contacts with only one monomer.
These particles can rotate with no energy cost because of this contact placement, so their rotational DOFs must be excluded in the definition of the isostatic contact number.
We inspected 3D dimer packings and found that there also exists this type of configuration.
Therefore, we employed the definition of the isostatic contact number for 3D dimers of
\begin{align}
 z^N_\text{iso} = 2d_f - \frac{2\pqty{d + N_\text{rr}}}{N}, \label{zNiso}
\end{align}
where $d_f \equiv 5$ is the number of DOFs, $d \equiv 3$ is the spatial dimension, and $N_\text{rr}$ is the number of rotational DOFs that are not constrained in the rotational rattler configurations.
Practically, we used the number of zero-frequency modes instead of $d + N_\text{rr}$ to calculate $z^N_\text{iso}$.
We counted the contact number of 3D dimer packings for each $\Delta\varphi$, and obtained that the excess contact number $\Delta z = z - z^N_\text{iso}$ is proportional to the square root of the excess density $\Delta\varphi$ (Fig.~\ref{fig:dz}),
\begin{align}
 \Delta z = C_z\pqty{\alpha}\Delta\varphi^{1/2},
\end{align}
and this relation holds for different aspect ratios.
When the system is near the jamming transition, the finite-size effect is observed.\cite{Goodrich_2012}
However, in the range of $\Delta\varphi \leq \num{e-6}$, approximately \SIrange[range-phrase = --]{10}{60}{\percent} of the packings have the same contact number of $z^N_\text{iso}$, defined as in~\eqref{zNiso}.
In other words, we could not observe the one additional contact attributed to a positive bulk modulus.\cite{Goodrich_2012}
We excluded the data from those packings in Fig.~\ref{fig:dz}.
The coefficient $C_z\pqty{\alpha}$ is nearly constant for each aspect ratio.

Similar to 2D dimers, our model of 3D dimers has certain types of contact that cannot be addressed: ``double'' and ``cusp'' contacts.
See Ref.~\onlinecite{Shiraishi_2019} for the definition of these contacts.
We searched 3D dimer packings for these contacts and identified the density range without these contacts as $\Delta\varphi \leq \num{e-2}$ for $\alpha = \num{1.2}$ and $\Delta\varphi \leq \num{e-1}$ for $\alpha = \num{1.5}$ and $\alpha = \num{2.0}$ (data not shown).
Here and hereafter, we excluded the results obtained from the configurations with densities above this range.

\subsection{Mechanical properties}\label{sec:mechprop}
In this section, we discuss the pressure, bulk modulus, and shear modulus of 3D dimer packings in terms of their $\Delta\varphi$ dependence.
We note that the stress tensor of granular assemblies is given by the Love-Weber formula.\cite{Weber_1966,Christoffersen_1981,Nicot_2013}
Anisotropic particles, including dimers, are modeled by granular assemblies, so the exact expression of the stress tensor is given by this formula.
However, previous work~\cite{Schreck_2010} has demonstrated that for the case of dimers, the virial expression~\cite{Allen_Tildesley} gives quantitatively similar values of the stress tensor as the Love-Weber formula.
Therefore, we employ the virial expression in this work.

\subsubsection{Pressure}
\begin{figure}[tb]
 \centering
 \includegraphics[width=\linewidth]{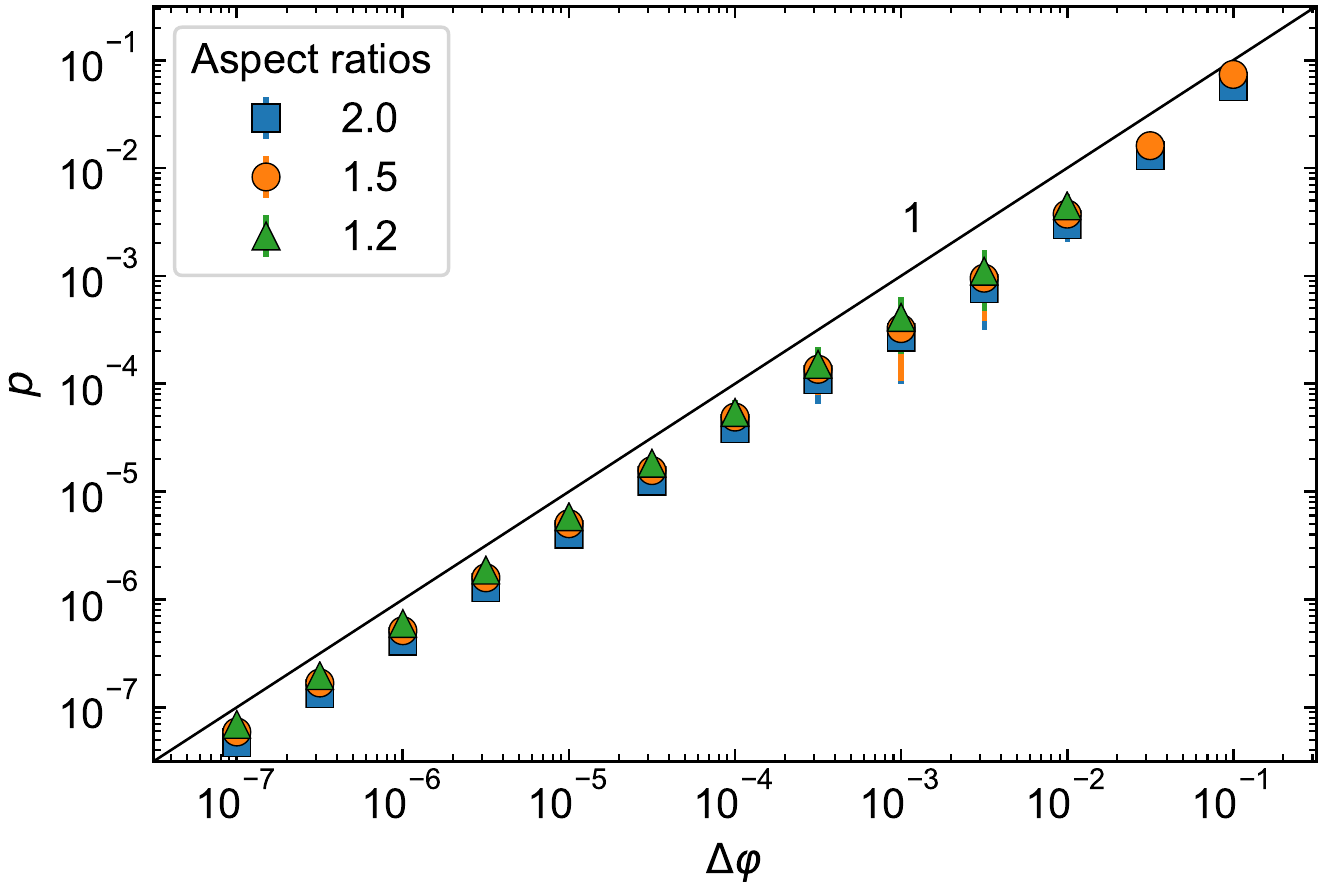}
 \caption{Pressure of 3D dimers with $\alpha=\num{1.2},\num{1.5},\num{2.0}$ as a function of excess density $\Delta\varphi$.}
 \label{fig:pressure}
\end{figure}

The stress tensor of 3D dimers is given in the virial expression as~\cite{Allen_Tildesley}
\begin{align}
 \sigma_{\alpha\beta} = -\frac{1}{2L^d}\sum_{ij}\pqty{F_{ij}^\alpha \rij^\beta + F_{ij}^\beta \rij^\alpha},
\end{align}
where $r_{ij}^\alpha$ is the $\alpha$-component of the vector between the centers of dimer $i$ and dimer $j$, and $F_{ij}^\alpha$ is the $\alpha$-component of the interparticle force between the two dimers:
\begin{align}
 F_{ij}^\alpha = \sum_{n_i n_j}\pqty{-\pdv{V}{\rkikj{i}{j}}} \frac{\rkikj{i}{j}^\alpha}{\rkikj{i}{j}},
\end{align}
where $\alpha,\beta = x,y,z$.
The pressure tensor is $p_{\alpha\beta} = -\sigma_{\alpha\beta}$, and the pressure is calculated as $p = \sum_\alpha p_{\alpha\alpha}/d$.
Figure~\ref{fig:pressure} shows that the pressure is proportional to $\Delta\varphi$, which is consistent with the results of spherical particles:\cite{O_Hern_2003}
\begin{align}
 p = C_p\pqty{\alpha}\Delta\varphi.
\end{align}
The coefficient $C_p\pqty{\alpha}$ weakly depends on the aspect ratio.

\subsubsection{Bulk modulus}
\begin{figure}[tb]
 \centering
 \includegraphics[width=\linewidth]{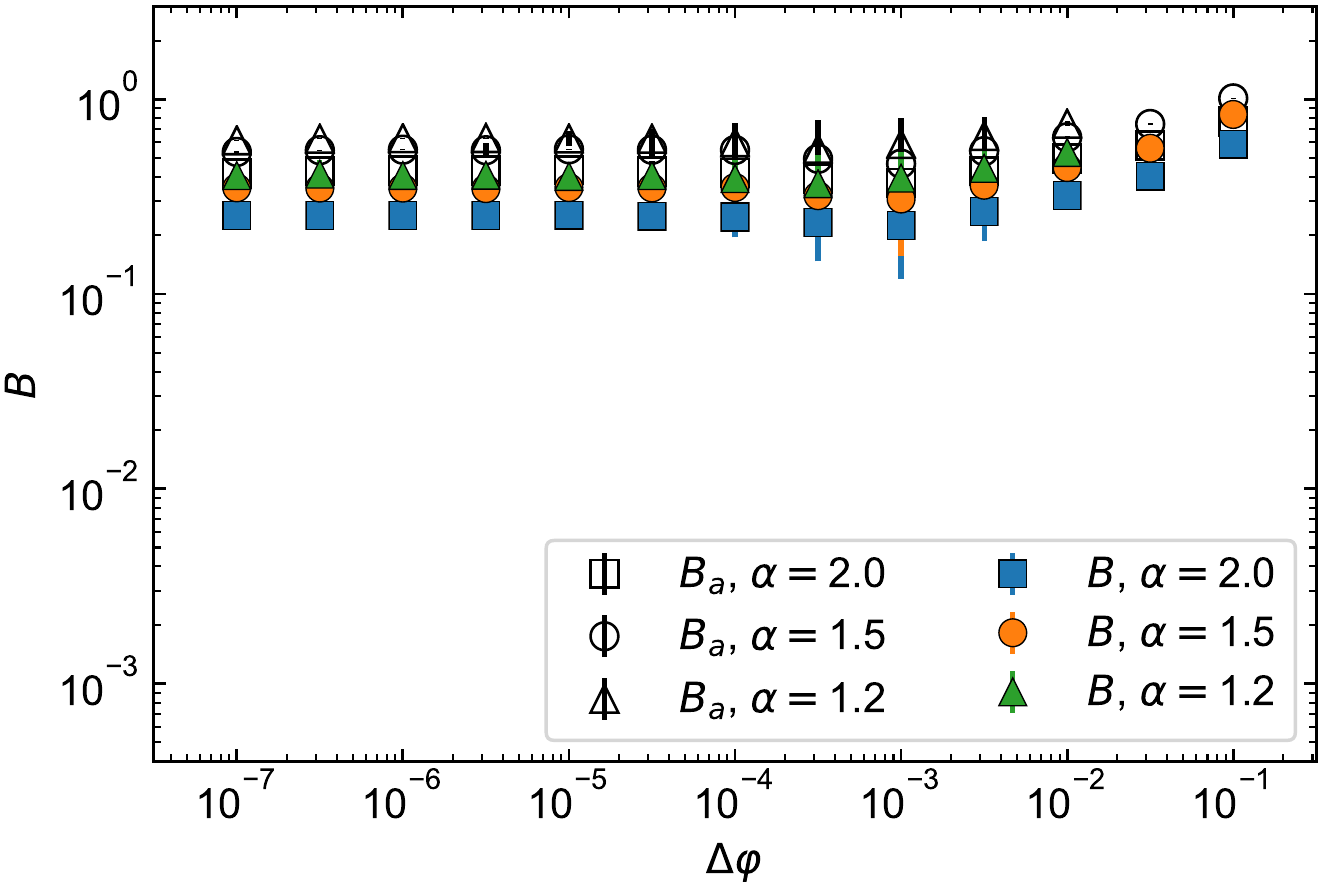}
 \caption{Bulk modulus $B$ and affine bulk modulus $B_a$ of 3D dimers with $\alpha=\num{1.2},\num{1.5},\num{2.0}$ as functions of excess density $\Delta\varphi$.}
 \label{fig:bulkmod}
\end{figure}

Next, we calculated the bulk modulus which is given by $B = \varphi\dd{p}/\dd{\varphi}$.
We first applied an affine bulk deformation to the packing ($\dd{\varphi} = \num{e-8}$) and calculated the pressure.
The bulk modulus calculated by using this pressure is referred to as the affine bulk modulus $B_a$.
The dimer packings are disordered, so subsequent relaxation motions occur.
We conducted energy minimization on the deformed packing and calculated the pressure.
The bulk modulus calculated by using the pressure measured after the energy minimization is referred to as the bulk modulus $B$.
The two bulk moduli as a function of $\Delta\varphi$ are shown in Fig.~\ref{fig:bulkmod}:
\begin{align}
 B \sim \Delta\varphi^0.
\end{align}
The bulk modulus is smaller than the affine bulk modulus because of the relaxation.
However, the bulk modulus remains as a finite value when the packings approach the jamming.
These behaviors are similar to those of spherical particles.\cite{O_Hern_2003}

\subsubsection{Shear modulus}
\begin{figure}[tb]
\centering
 \includegraphics[width=\linewidth]{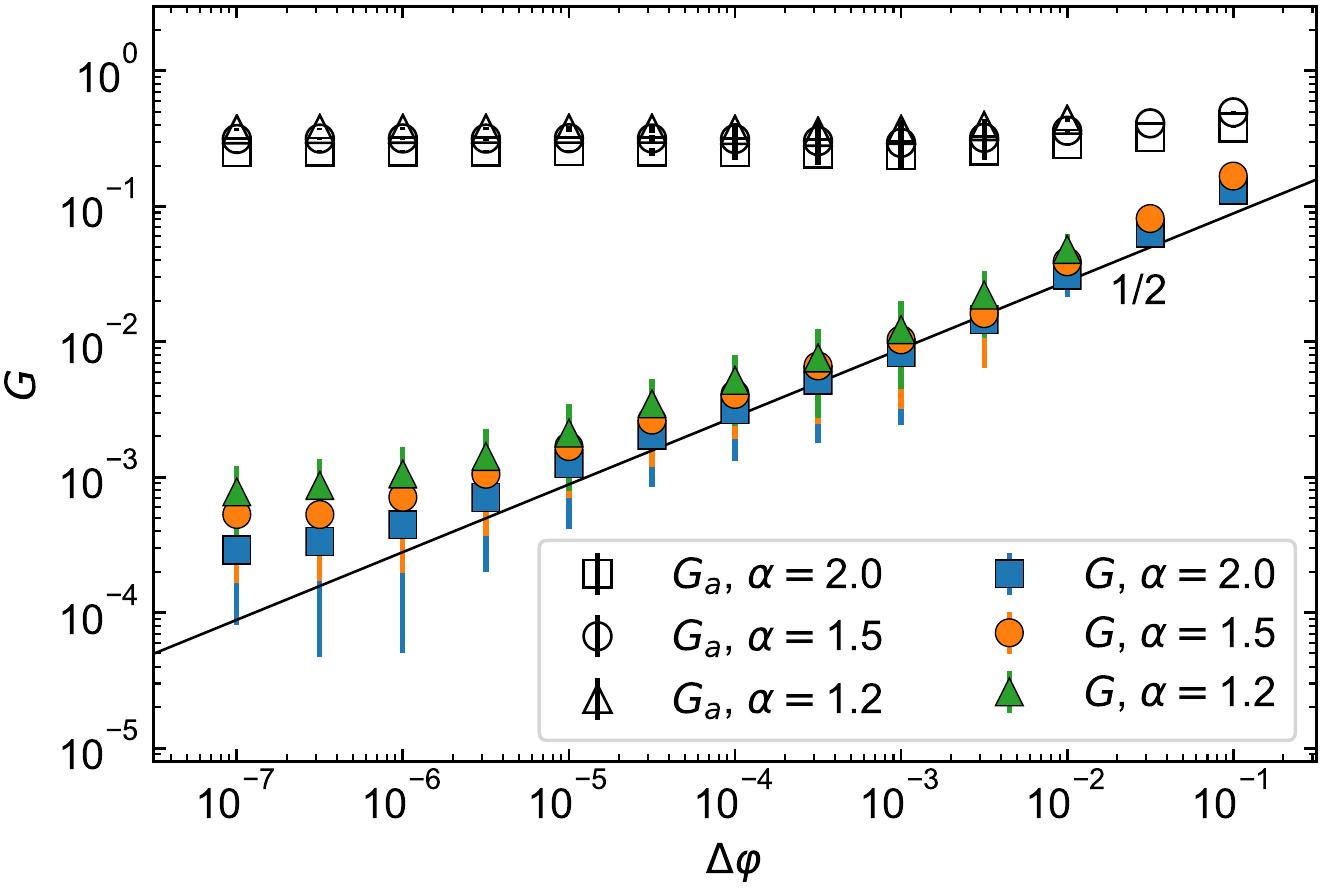}
 \caption{Shear modulus $G$ and affine shear modulus $G_a$ of 3D dimers with $\alpha=\num{1.2},\num{1.5},\num{2.0}$ as functions of excess density $\Delta\varphi$.}
 \label{fig:shearmod}
\end{figure}

Here, we calculated the shear modulus by applying a shear strain $\gamma$ in the $x$ direction with a strain gradient in the $y$ direction.
To do so, we must determine the affine shear strain on the rotational DOFs of 3D dimer particle.
The displacements of the translational DOFs are trivial.
For the rotational DOFs, we first consider the deformation gradient tensor $F$ for the shear strain
\begin{align}
 F = \pmqty{1 & \gamma & 0 \\ 0 & 1 & 0 \\ 0 & 0 & 1}.
\end{align}
To determine the rotational DOFs in a sheared configuration $\phi_i^\prime$ and $\theta_i^\prime$, we consider a normalized orientational vector of a 3D dimer:
\begin{align}
 \vb{r}\pqty{\phi_i, \theta_i} = \pmqty{\sin\theta_i\sin\phi_i \\ -\sin\theta_i\cos\phi_i \\ \cos\theta_i}.
\end{align}
This vector is deformed by deformation gradient tensor $F$.
Thus, we solve the equation
\begin{align}
 \vb{r}\pqty{\phi_i^\prime, \theta_i^\prime} = \frac{F\vb{r}\pqty{\phi_i, \theta_i}}{\abs{F\vb{r}\pqty{\phi_i, \theta_i}}}
\end{align}
and obtain the affine shear deformation of 3D dimers as
\begin{align}
\begin{aligned}
 x_i^\prime      &= x_i + \gamma y_i, \quad y_i^\prime = y_i, \quad z_i^\prime = z_i,\\
 \phi_i^\prime   &= \tan^{-1}\pqty{\tan\phi_i - \gamma},\\
 \theta_i^\prime &= \tan^{-1}\pqty{\tan\theta_i \frac{\cos\phi_i}{\cos\phi_i^\prime}}. \label{affineshear}
\end{aligned}
\end{align}

The shear modulus is obtained by $G = \dd{\sigma_{xy}}/\dd{\gamma}$.
Similar to the bulk modulus, we use the affine shear modulus $G_a$ to refer to the modulus measured with the stress calculated immediately after the affine strain~\eqref{affineshear} was applied.
In contrast, we use the shear modulus $G$ to refer to the modulus measured with the stress calculated after the energy minimization was performed.
The results are shown in Fig.~\ref{fig:shearmod}.
The power-law scaling
\begin{align}
 G = C_G\pqty{\alpha} \Delta\varphi^{1/2}
\end{align}
was observed, where the $\Delta\varphi$ dependence is identical for spherical particles~\cite{O_Hern_2003} and 2D dimers~\cite{Schreck_2010}.
$G$ develops a plateau at $\Delta\varphi=\num{e-6.5}$, where we consider the finite-size effect as observed in spherical particles.\cite{Goodrich_2012}
The coefficient $C_G\pqty{\alpha}$ shows a weak dependence on $\alpha$.

\subsection{Vibrational properties}
In this section, we discuss the properties of vibrational eigenstates, which are obtained by diagonalizing the dynamical matrix~\eqref{dynmat_def}.

We calculated the vibrational density of states using the set of eigenfrequencies $\omega_k$ as
\begin{align}
 g\pqty{\omega} = \frac{1}{d_fN - d - N_\text{rr}} \sum^{d_fN - d - N_\text{rr}}_{k=1} \delta\pqty{\omega - \omega_k},
\end{align}
where $\delta\pqty{x}$ is the Dirac delta function.
We also calculated the participation ratio~\cite{Dean_1972} of each eigenmode $\vb{e}_k$
\begin{align}
 p_k = \frac{\pqty{\sum^N_{i=1} \abs{\vb{e}_k^i}^2}^2}{N\sum^N_{i=1}\abs{\vb{e}_k^i}^4}
\end{align}
to study the extent of spatial localization of mode $k$.
If $p_k = 1$, then all particles in the system participate in mode $k$.
If $p_k = 1/N$, then only one particle in the system participates in mode $k$.
The eigenmodes consist of translational and rotational elements.
Following the study on ellipsoidal particles,\cite{Mailman_2009} we defined the quantity $R_k$ to evaluate the contribution of rotational motions as
\begin{align}
 R_k = 1 - \sum^N_{i=1} \sum_{d = x,y,z} \pqty{e_k^{i,d}}^2,
\end{align}
and calculated $R_k$ for each mode.
If $R_k = 1$, then mode $k$ consists of only rotational motions.
If $R_k = 0$, then only translational motions compose mode $k$.

We also calculated the participation ratio and contribution of rotations based on the real-space displacements.\cite{Shiraishi_2019}
The translational displacement field of each monomer was used to express the vibrational eigenstates.
We denote the translational displacements for the two monomers of dimer $i$ as $\tilde{\vb{e}}_k^i$ (size $2d$) and the monomer displacements of the entire packing as $\tilde{\vb{e}}_k$ (size $2dN$).
The elements are written as
\begin{align}
 \begin{aligned}
  \tilde{\vb{e}}_k   &= \pmqty{\tilde{\vb{e}}_k^1 & \cdots & \tilde{\vb{e}}_k^N}, \\
  \tilde{\vb{e}}_k^i &= \pmqty{\tilde{e}_k^{\pqty{i,+1},x} & \tilde{e}_k^{\pqty{i,+1},y} & \tilde{e}_k^{\pqty{i,+1},z} & \tilde{e}_k^{\pqty{i,-1},x} & \tilde{e}_k^{\pqty{i,-1},y} & \tilde{e}_k^{\pqty{i,-1},z}},
 \end{aligned}
\end{align}
where
\begin{widetext}
\begin{align}
 \begin{aligned}
  \tilde{e}_k^{\pqty{i,\pm 1},x} &= e_k^{i,x} \pm \frac{b\pqty{\alpha-1}}{2}\bqty{\sin\pqty{\frac{\tilde{\theta}_i+e_k^{i,\tilde{\theta}}}{\sqrt{I_1}}} \sin\pqty{\frac{\tilde{\phi}_i+e_k^{i,\tilde{\phi}}}{\sqrt{I_1}\sin\theta_i}} - \sin\frac{\tilde{\theta}_i}{\sqrt{I_1}}\sin\frac{\tilde{\phi}_i}{\sqrt{I_1}\sin\theta_i}}, \\
  \tilde{e}_k^{\pqty{i,\pm 1},y} &= e_k^{i,y} \pm \frac{b\pqty{\alpha-1}}{2}\bqty{-\sin\pqty{\frac{\tilde{\theta}_i+e_k^{i,\tilde{\theta}}}{\sqrt{I_1}}} \cos\pqty{\frac{\tilde{\phi}_i+e_k^{i,\tilde{\phi}}}{\sqrt{I_1}\sin\theta_i}} + \sin\frac{\tilde{\theta}_i}{\sqrt{I_1}}\cos\frac{\tilde{\phi}_i}{\sqrt{I_1}\sin\theta_i}}, \\
  \tilde{e}_k^{\pqty{i,\pm 1},z} &= e_k^{i,z} \pm \frac{b\pqty{\alpha-1}}{2}\bqty{\cos\pqty{\frac{\tilde{\theta}_i+e_k^{i,\tilde{\theta}}}{\sqrt{I_1}}} - \cos\frac{\tilde{\theta}_i}{\sqrt{I_1}}}.
 \end{aligned}
\end{align}
\end{widetext}
Using this real-space eigenvector, we defined the participation ratio and contribution of rotations based on the real-space displacements as
\begin{align}
 \tilde{p}_k &= \frac{\pqty{\sum_{i=1}^{2N}\abs{\tilde{\vb{e}}_k^i}^2}^2}{2N\sum_{i=1}^{2N}\abs{\tilde{\vb{e}}_k^i}^4}, \\
 \tilde{R}_k &= \frac{\abs{\tilde{\vb{e}}_k - \tilde{\vb{e}}_k^\text{trans}}}{\abs{\tilde{\vb{e}}_k}}.
\end{align}
Here, $\tilde{\vb{e}}_k^\text{trans}$ (size $2dN$) is defined as the components of translational DOFs in the original $\vb{e}_k$ (size $d_fN$).

In the following (Sect.~\ref{sec:unstressed}), we compare the properties of the original system to those of the unstressed system.
The analysis of the unstressed system of dimers corresponds to all stiffnesses being retained between monomers but all forces being dropped in the dynamical matrix and execution of the same calculation.
Since we modeled the interaction potential between monomers as purely repulsive interactions, we refer to the original state as the stressed system.
We calculated the eigenfrequencies and eigenvectors of the unstressed system and obtained the VDOS $g\pqty{\omega}$, participation ratio $p_k$, and contribution of rotations $R_k$.
Please refer to Sect.~\ref{sec:method_unstress} for the unstressed system.

We finally changed the excess density of the packings in the stressed system and observed a critical behavior of a characteristic frequency of the VDOS.

\subsubsection{General description of vibrational properties}
\begin{figure}[tb]
 \centering
 \includegraphics[width=\linewidth]{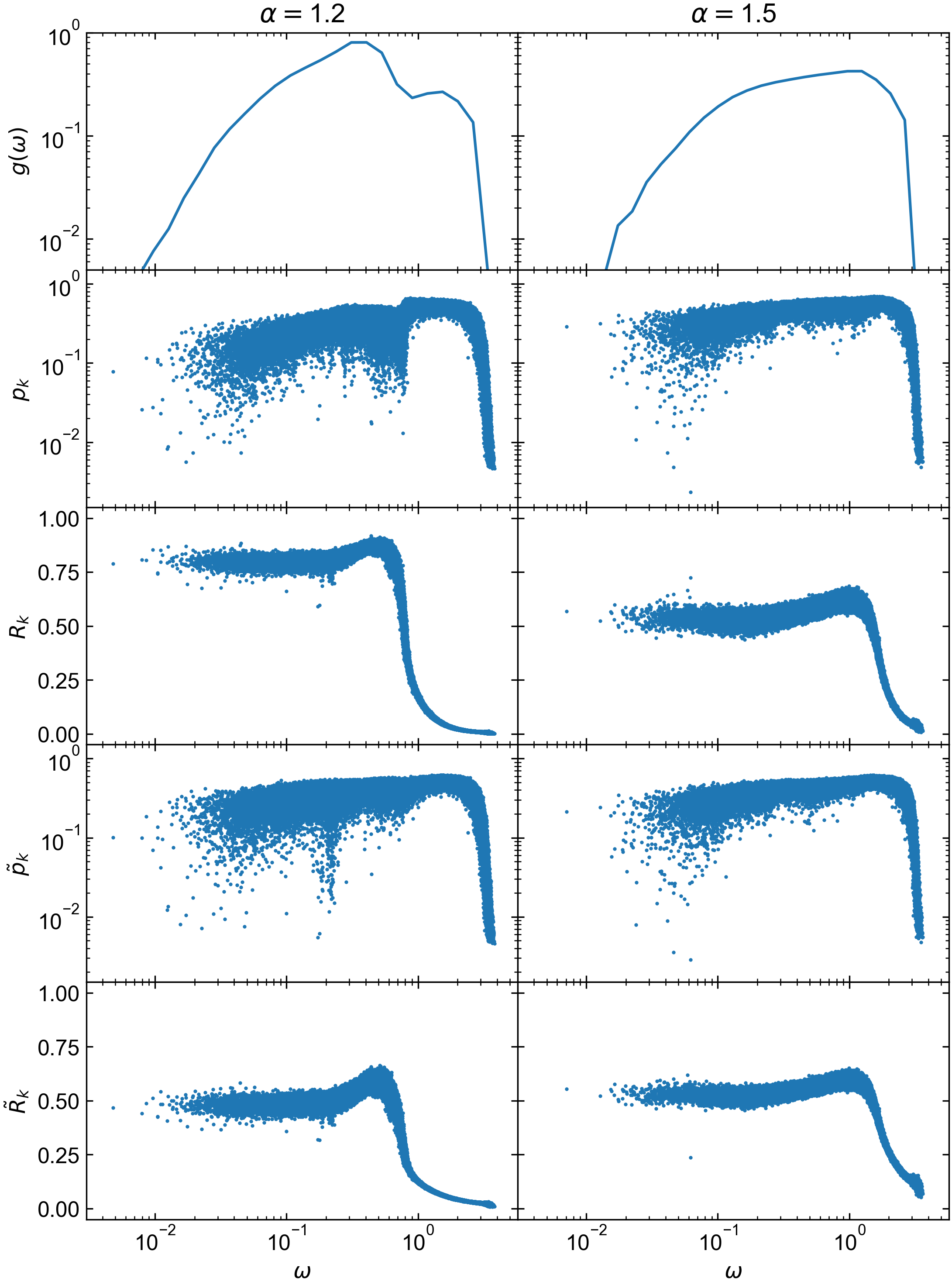}
 \caption{The VDOS $g\pqty{\omega}$, participation ratio $p_k$, contributions of rational motions $R_k$, and their real-space interpretations $\tilde{p}_k$ and $\tilde{R}_k$ of 3D dimers with $\alpha = \num{1.2}$ (left row) and $\alpha = \num{1.5}$ (right row). The excess density of the packings is $\Delta\varphi = \num{e-2}$ for both cases.}
 \label{fig:stressed}
\end{figure}

We summarize the vibrational properties of 3D dimers with $\alpha=\num{1.2}$ in the panels in the left row of Fig.~\ref{fig:stressed}.
The top panel shows the VDOS $g\pqty{\omega}$.
It exhibits a peak at $\omega \simeq 0.3$, and there is a plateau above this frequency and a shoulder below this frequency.
The third panel shows $R_k$, which is the contribution of the rotational motions to each mode.
It is nearly zero for $\omega \gtrsim 0.3$, increases at $\omega \approx 0.3$, and remains large for $\omega \lesssim 0.3$.
Therefore, the plateau in the higher-frequency region mainly consists of translational modes, whereas the peak and shoulder in the lower-frequency region consist of rotational modes.
We shall refer to the high-frequency plateau as the translational plateau and the low-frequency shoulder as the rotational plateau, as it becomes a plateau near the jamming transition.
The second panel shows the participation ratio $p_k$.
Clearly, the modes in the highest-frequency edge have a strongly localized nature.
In the translational plateau, $p_k$ has a larger value: $p_k \simeq \num{0.6}$; therefore, the modes are extended.
At the peak and in the rotational plateau, $p_k$ slightly decreases but remains at $p_k \gtrsim \num{0.1}$.
In the lowest-frequency region, however, there appear several spatially localized modes.
This is reminiscent of the quasilocalized modes of the random packing of spherical particles, although our system size is too small to study the quantitative properties of these modes.
The real-space versions of $\tilde{R}_k$ and $\tilde{p}_k$ exhibit qualitatively similar behaviors to $R_k$ and $p_k$, respectively~\footnote{%
However, we note two quantitative differences; $\tilde{p}_k$ does not show a slight decrease in the peak and the rotational plateau, and $\tilde{R}_k$ are slightly smaller than $R_k$  at the peak and in the rotational plateau.}.

Next, we consider the dimers with $\alpha=\num{1.5}$ (the right row in Fig.~\ref{fig:stressed}).
The distinct peak of $g\pqty{\omega}$ in the case of $\alpha=\num{1.2}$ disappears for $\alpha=\num{1.5}$.
$R_k$ is nearly zero at the highest frequency, and it increases as the frequency decreases.
The data of $R_k$ suggest that the rotational modes of $\alpha = \num{1.5}$ dimers shift to higher frequency than those of the $\alpha = \num{1.2}$ dimers, and they mix with the translational modes.
The $p_k$ of $\alpha = \num{1.5}$ behaves similarly as that of the $\alpha = \num{1.2}$ dimers.
The real-space versions of $\tilde{R}_k$ and $\tilde{p}_k$ show no significant differences with $R_k$ and $p_k$.
We conducted the same calculation for dimers with $\alpha = \num{2.0}$.
The peak of $R_k$ slightly shifts to a higher frequency than in the case of $\alpha=\num{1.5}$, but all other features described above are valid in the case of $\alpha = \num{2.0}$.
Therefore, the separation of the translational and rotational modes becomes weaker with increasing aspect ratio.

\subsubsection{Comparison with the unstressed system}\label{sec:unstressed}
In this subsection, we discuss the role of stress in the vibrational properties of 3D dimers.
To this end, we study the vibrational properties of the unstressed system, where all contacts in the original stressed system are replaced with relaxed springs (see Sect.~\ref{sec:method_unstress}).
For sphere packings, the stress does not play an important role in the vibrational modes in the plateau; namely, the stressed and unstressed systems exhibit similar plateaus and similar critical behavior near the jamming transition.
However, the stress is known to make the system mechanically unstable,\cite{Wyart_PRE_2005,Lerner_2014} which results in an abundance of very low-frequency vibrational modes.\cite{Wyart_PRE_2005,Lerner_2014}
In particular, quasilocalized vibrational modes are induced in the very low-frequency region by the stress.\cite{Mizuno_2017,Lerner_2018,Shimada_Mizuno_Wyart_Ikeda_2018}
By contrast, the stress is known to stabilize the lowest-frequency vibrations in the random packings of ellipsoidal particles.\cite{Mailman_2009,Schreck_2012}

\begin{figure}[tb]
 \centering
 \includegraphics[width=\linewidth]{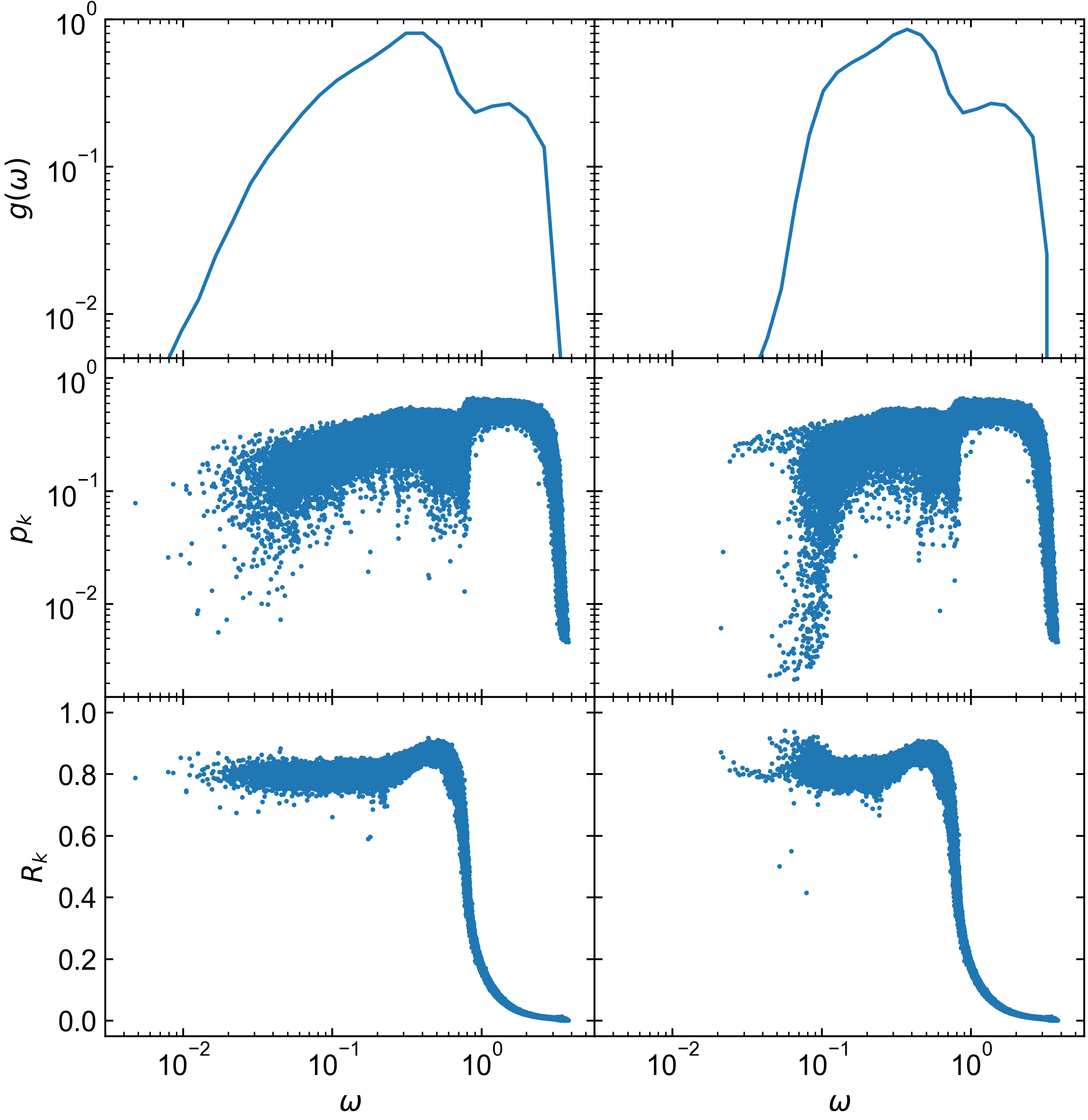}
 \caption{Comparison of $g\pqty{\omega}$, $p_k$, and $R_k$ between the stressed system (left row) and the unstressed system (right row) of 3D dimer packings with $\alpha=\num{1.2}$ and $\Delta\varphi=\num{e-2}$.}
 \label{fig:vs_unst2}
\end{figure}

\begin{figure}[tb]
 \centering
 \includegraphics[width=\linewidth]{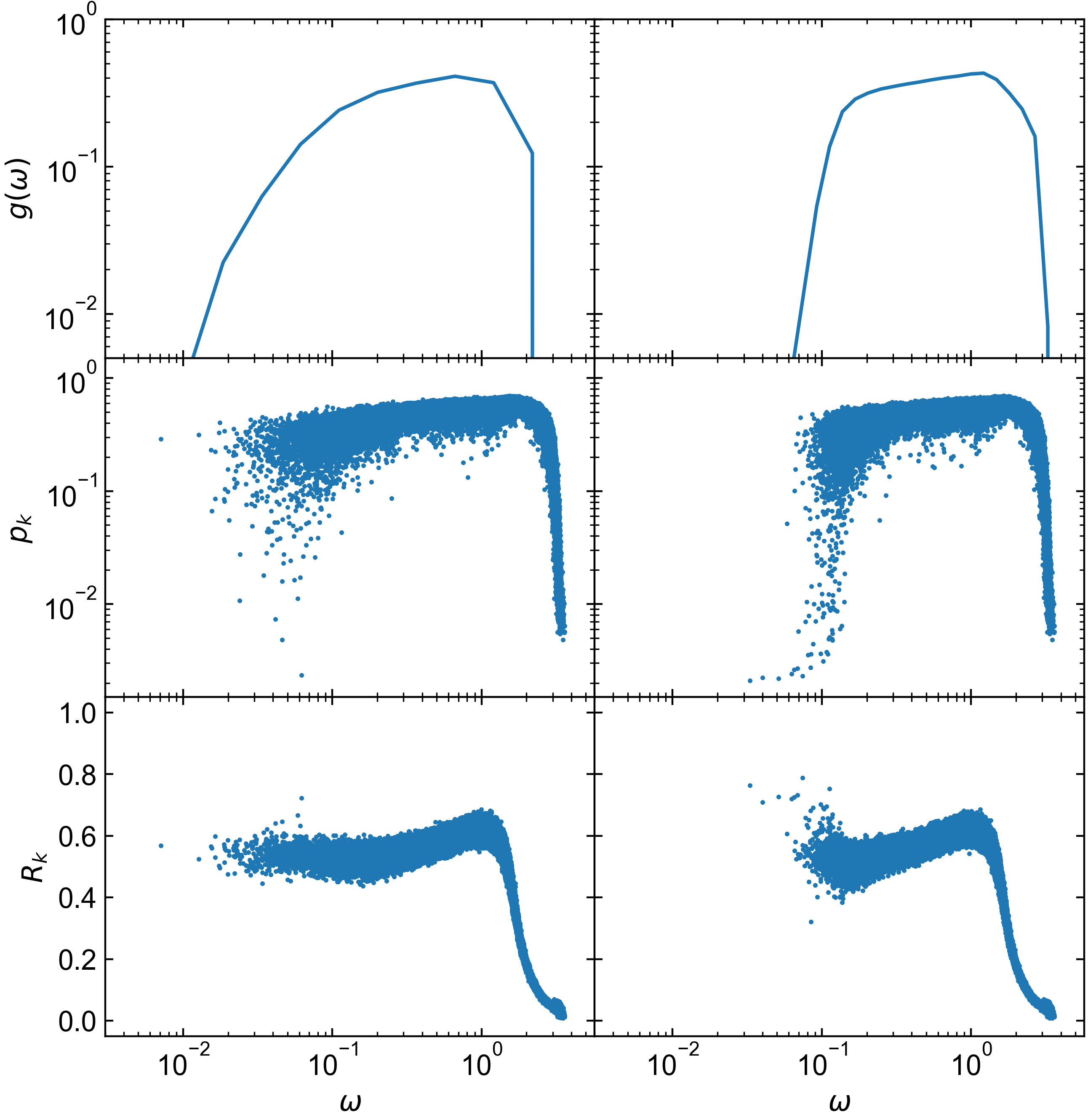}
 \caption{Comparison of $g\pqty{\omega}$, $p_k$, and $R_k$ between the stressed system (left row) and the unstressed system (right row) of 3D dimer packings with $\alpha=\num{1.5}$ and $\Delta\varphi=\num{e-2}$.}
 \label{fig:vs_unst5}
\end{figure}

The VDOS $g\pqty{\omega}$, participation ratio $p_k$, and contribution of rotations $R_k$ of the unstressed system of dimers with $\alpha=\num{1.2}$ and $\Delta\varphi=\num{e-2}$ are shown in Fig.~\ref{fig:vs_unst2}.
Clearly, all of these quantities of the unstressed system are identical to those of the stressed system in the translational plateau; thus, the stress does not play an important role in the translational modes in the plateau.
The peak at $\omega \approx 0.3$ is also not affected by the stress.
However, the differences between the stressed and unstressed systems are clear in the low-frequency region.
The VDOS $g\pqty{\omega}$ of the unstressed system forms a clearer plateau and sharply drops below the plateau.
The $R_k$ value of the unstressed and stressed systems are similar in this frequency region, i.e., the modes remain to have a rotational nature.
These results imply that the stress induces an abundance of low-frequency rotational vibrational modes.
This is similar to the case of spheres, except that the low-frequency vibrations are rotational in dimers.
The participation ratio of the unstressed system shows a drop at the onset frequency of the plateau, where the modes below the plateau have higher participation ratios.
These results are similar to the results of spheres;\cite{Mizuno_2017} however, the drop at the onset of the plateau is much sharper in dimers.
To elucidate the quasilocalized nature of the very low-frequency vibrations in the stressed and unstressed systems, much larger systems are required, which can be addressed in the future.

The results of the unstressed system of the dimers with $\alpha=\num{1.5}$ and $\Delta\varphi=\num{e-2}$ are shown in Fig.~\ref{fig:vs_unst5}.
The impact of the stress in the $\alpha=\num{1.5}$ case is almost the same as that in $\alpha=\num{1.2}$.
The VDOS $g\pqty{\omega}$ above the plateau does not change in the unstressed system, and the participation ratio and contribution of rotations do not change either.
Below the plateau, the VDOS $g\pqty{\omega}$ sharply drops in the unstressed system.
The number of modes with lower participation ratio increases in the unstressed system.
The contribution of rotations $R_k$ has basically the same properties in the unstressed system.

\subsubsection{Critical behavior near the jamming transition}\label{sec:omegastar}
\begin{figure}[tb]
 \centering
 \includegraphics[width=\linewidth]{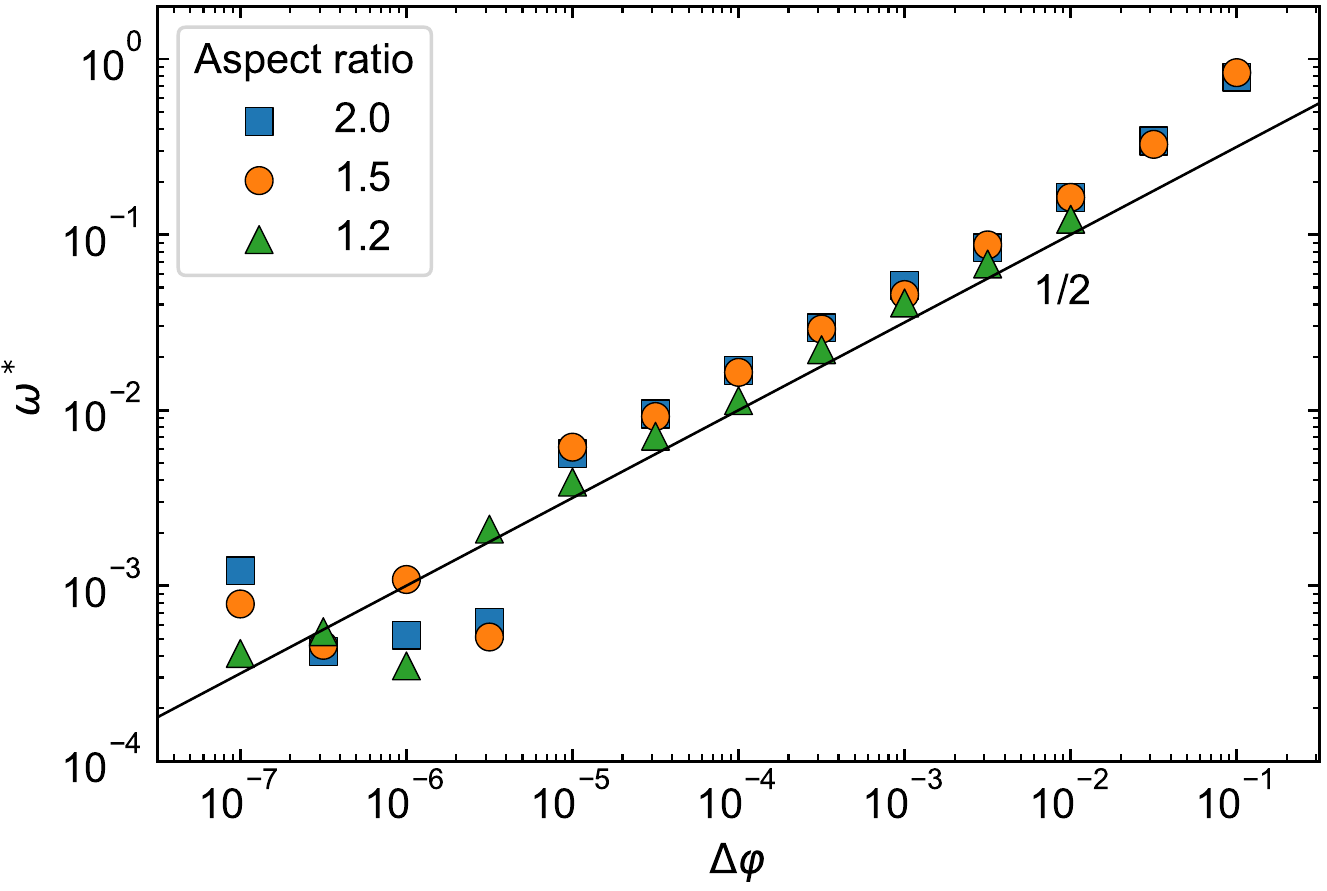}
 \caption{Onset frequency of the rotational plateau $\omega^*$ of 3D dimers with $\alpha=\num{1.2},\num{1.5},\num{2.0}$, which shows a power-law scaling $\omega^* \propto \Delta\varphi^{1/2}$. Due to the system size of our simulation, this scaling does not hold for the densities that are very close to the jamming transition ($\Delta\varphi \leq \num{e-5.5}$).}
 \label{fig:omegastar}
\end{figure}

We finally changed the distance from the jamming transition and observed the behavior of the VDOS.
When approaching the jamming transition, the rotational plateau extends toward zero frequency, similar to 2D dimers.\cite{Shiraishi_2019}
We defined the onset frequency of the rotational plateau in the unstressed system as $\omega^*$.
Practically, $\omega^*$ was calculated as the peak frequency of $g\pqty{\omega}/\omega$ of the unstressed system.
We conducted the calculation for the dimers with $\alpha=\num{1.2},\num{1.5},\num{2.0}$, and the dependence on $\Delta\varphi$ is shown in Fig.~\ref{fig:omegastar}.
The $\omega^*$ of each aspect ratio is proportional to the square root of the excess density, so it is proportional to the excess contact number:
\begin{align}
 \omega^* = C_\omega\pqty{\alpha}\Delta\varphi^{1/2} \propto \Delta z.
\end{align}
This result implies that the contact number controls the onset frequency of the plateau in the same way as for spherical particles.\cite{Wyart_EPL_2010}
We have confirmed that $\omega^*$, defined as the onset frequency of the plateau of the unstressed system, corresponds to the frequency where the VDOS $g\pqty{\omega}$ of the unstressed system deviates from that of the stressed system.

\section{Conclusion}
In this article, we revealed the mechanical and vibrational properties of random packings composed of 3D dimer particles.
First, we formulated the equations of motion of the system and linearized them to obtain the dynamical matrix.
This discussion is valid for symmetrical-top-type nonspherical particles such as ellipsoids or spherocylinders.

We generated 3D dimer packings at the jamming transition and calculated the jamming density for various aspect ratios.
Our results were comparable to the theoretical prediction by Baule et al.~\cite{Baule_2013}, and especially there are quantitative agreement for the value and location of the maximum density.
The contact numbers and coordination numbers of the packings at the jamming transition were also calculated.
The results on the contact number indicate that the 3D dimers are isostatic, and those of the coordination number suggests the similarity between the particle coordination of dimers and spherocylinders.

We continued the compression and obtained packings with density higher than that at the jamming transition.
The contact number of the packings was calculated, and it was found that the excess contact number was proportional to the square root of the excess density.
The mechanical properties, that is, the pressure, bulk modulus, and shear modulus, of the packings were also calculated as a function of the excess density.
All three quantities exhibited the same $\Delta\varphi$ dependencies as spherical particles.
Especially, the non-affine relaxation was dominant in the response to shear deformation, so the shear modulus was proportional to the square root of the excess density.

We finally examined the vibrational properties of 3D dimer packings.
The VDOS of 3D dimers has two plateaus, and a peak divides these plateaus.
By calculating the rotational contribution of the modes $R_k$, we concluded that rotational components are dominant in the modes that constitute the peak and lower-frequency plateau.
We also calculated the participation ratio and found that the modes in the two plateaus have a spatially extended nature.
In the frequency region below these plateaus, some modes show a quasilocalized nature.
In addition to the original system, we also analyzed the unstressed system of the packings.
The properties of the modes above the lower-frequency rotational plateau are identical between the stressed and unstressed systems, which indicates that the stress does not play an important role in the high-frequency modes.
In the low-frequency region, the VDOS of the unstressed system is strongly suppressed compared with the stressed system.
In other words, in the case of spheres, the stress induces an abundance of low-frequency vibrations in the dimers.
However, in the unstressed dimers, we observed a very sharp drop in the participation ratio near the onset of the plateau.
When approaching the jamming transition, the low-frequency rotational plateau extends towards zero frequency.
The onset frequency $\omega^*$ of the rotational plateau shows a critical scaling $\omega^* \propto \Delta z$, which is consistent with the observation in 2D dimers.
Therefore, the plateau is controlled by the contact number as understood for spherical particles.

Our results show that the dimer particles share many commonalities with spherical particles in terms of isostaticity, non-affine relaxation, abundance of very low-frequency vibrations in the stressed system, and critical behavior of $\omega^*$, but the low-frequency vibrations are rotational in dimers.
This result markedly contrasts with the case of ellipsoids, which are not isostatic at the jamming but hypostatic,\cite{Donev_2004} i.e., the number of contacts is less than the number of DOFs.
The shear modulus of ellipsoidal packings shows a different scaling $G \propto \Delta\varphi$.\cite{Mailman_2009,Schreck_2010}
The VDOS of ellipsoidal packings shows the same critical scaling as spherical packings,\cite{Zeravcic_2009} but it possesses the unusual low-frequency modes caused by the hypostaticity.\cite{Mailman_2009}
The energy quartically increases due to the packing deformation along the modes.\cite{Mailman_2009,Schreck_2012}
The scaling of the shear modulus originates from the quartic modes.\cite{Mailman_2009}
The effect of the stresses in the dynamical matrix is different from that for spherical packings: these stresses stabilize the quartic modes.\cite{Schreck_2012,Harukuni_2018}
Further study is required to clarify the attributes that cause these discrepancies between the properties of dimers and those of ellipsoids.

Due to the limitation of the system size, the present work cannot access the very low-frequency region, where phonon modes and quasilocalized modes coexist in the random packings of spheres.\cite{Mizuno_2017}
According to our results, the rotational modes appear in a lower-frequency region than the translational modes.
Therefore, it is an interesting direction to elucidate the details of the quasilocalized nature of the low-frequency vibrational modes and their statistical properties in the random packings of dimers.

\begin{acknowledgments}
This work was supported by JSPS KAKENHI Grant Numbers 16H04034, 17H04853, 18H05225, 19H01812, 19K14670, and 20H01868.
This work was also supported by the Asahi Glass Foundation.
The computations were partially performed using the Research Center for Computational Science, Okazaki, Japan.
\end{acknowledgments}

\bibliography{bibliography.bib}

\newpage
\begin{widetext}
\begin{center}
\textbf{\large Supplemental Material for ``Mechanical and Vibrational Properties of Three-Dimensional Dimer Packings Near the Jamming Transition''}

\medskip

Kumpei Shiraishi\textsuperscript{1}, Hideyuki Mizuno\textsuperscript{1}, and Atsushi Ikeda\textsuperscript{1,2}

\smallskip

\textsuperscript{1}\textit{Graduate School of Arts and Sciences, University of Tokyo, Komaba, Tokyo 153-8902, Japan}

\textsuperscript{2}\textit{Research Center for Complex Systems Biology, Universal Biology Institute, University of Tokyo, Komaba, Tokyo 153-8902, Japan}
\end{center}

\setcounter{section}{0}
\setcounter{figure}{0}
\setcounter{equation}{0}
\section{Calculation of the volume and moment of inertia}
\begin{figure}[b]
 \centering
 \includegraphics[width=.35\linewidth]{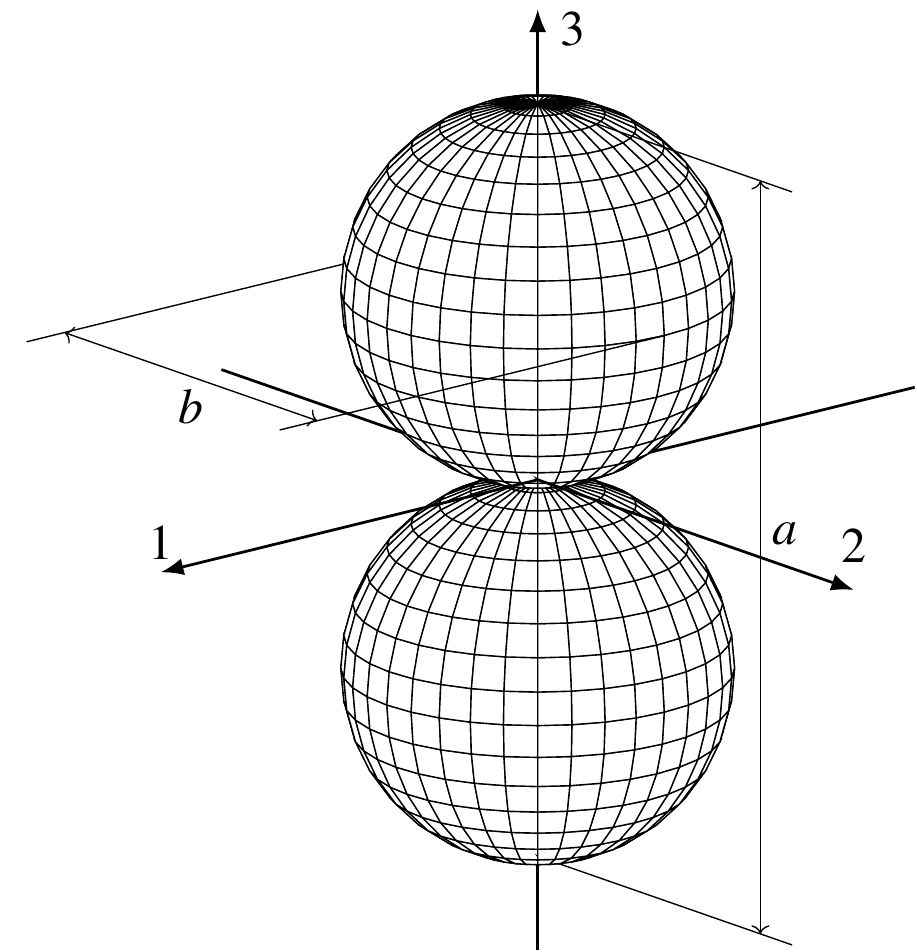}
 \caption{3D dimer particle of $\alpha=\num{2.0}$ on a coordinate system.}
 \label{fig:coord}
\end{figure}

In this section, we formulate the volume and principal moments of inertia of a 3D dimer.
Suppose that a 3D dimer is placed so that its major axis coincides with the $z$-axis of the coordinate system (c.f. Fig.~\ref{fig:coord}).
The length of the radius at $z$ is
\begin{align}
 R\pqty{z} = \sqrt{\pqty{\frac{b}{2}}^2 - \pqty{z - \frac{b\pqty{\alpha-1}}{2}}^2},
\end{align}
and the volume of the 3D dimer $v$ is given by
\begin{align}
 v = 2\pi\int^{b\alpha/2}_0\dd{z}R\pqty{z}^2
   = 2\pi\alpha^2\pqty{1 - \frac{\alpha}{3}}\pqty{\frac{b}{2}}^3.
\end{align}
We denote the mass density of the particle as $\rho = m/v$.

Here, we denote the principal axes of inertia as 1, 2, and 3, and place a 3D dimer as shown in Fig.~\ref{fig:coord}.
The major axis coincides with the 3-axis.
The principal moments of inertia are calculated as
\begin{align}
 I_i &= \int\rho\pqty{x_j^2 + x_k^2}\dd{v},
\end{align}
where $\pqty{i,j,k}$ represents $\pqty{1,2,3}$ and its cyclically permuted sets.
Because of the symmetry of axes 1 and 2, two of the principal moments of inertia are equal: $I_1 = I_2 \neq I_3$.
The expressions are given as
\begin{align}
\begin{aligned}
 I_1 &= \int \rho\pqty{{x_2}^2 + {x_3}^2}\dd{v} = \frac{m\alpha\pqty{\alpha^2 - 5\alpha + 20}}{20\pqty{3-\alpha}}\pqty{\frac{b}{2}}^2 = I_2, \\
 I_3 &= \int \rho\pqty{{x_1}^2 + {x_2}^2}\dd{v} = \frac{m\alpha\pqty{3\alpha^2 - 15\alpha + 20}}{10\pqty{3-\alpha}}\pqty{\frac{b}{2}}^2.
\end{aligned}
\end{align}
Therefore, 3D dimer is a symmetrical top.

\section{Dynamical matrix elements for 3D dimer packings}
The explicit elements of the dynamical matrix are given in this section.
By definition, this matrix is a real symmetric matrix.
The size of the matrix is $d_fN \times d_fN$, and the elements are denoted by $\Melem{\alpha}[\beta]{i}[j]$, where $\alpha,\beta = x,y,z,\tilde{\phi},\tilde{\theta}$ and $i,j = 1,\dots,N$.
In the following, we denote the force and stiffness that act on the two monomers as
\begin{align}
 \drv{i}{j}  &\equiv \frac{\epsilon}{\sij}\pqty{1 - \frac{\rkikj{i}{j}}{\sij}} H\pqty{1 - \frac{\rkikj{i}{j}}{\sij}}, \\
 \ddrv{i}{j} &\equiv \frac{\epsilon}{\sij^2} H\pqty{1 - \frac{\rkikj{i}{j}}{\sij}},
\end{align}
respectively.
We enumerate all cases by separating the diagonal blocks $i = j$ and the off-diagonal blocks $i \neq j$.

\subsection{Diagonal blocks}
\begin{align}
 \Melem{x}{a} = \sum_{i \neq a}^N\sum_{n_a n_i} \kf{a}{i} \frac{\xkikj{a}{i}^2}{\rkikj{a}{i}^2} - \frac{\drv{a}{i}}{\rkikj{a}{i}}
\end{align}
\begin{align}
 \Melem{y}{a} = \sum_{i \neq a}^N\sum_{n_a n_i} \kf{a}{i} \frac{\ykikj{a}{i}^2}{\rkikj{a}{i}^2} - \frac{\drv{a}{i}}{\rkikj{a}{i}}
\end{align}
\begin{align}
 \Melem{z}{a} = \sum_{i \neq a}^N\sum_{n_a n_i} \kf{a}{i} \frac{\zkikj{a}{i}^2}{\rkikj{a}{i}^2} - \frac{\drv{a}{i}}{\rkikj{a}{i}}
\end{align}
\begin{align}
\begin{aligned}
 \Melem{\tilde{\theta}}{a} = \sum_{i \neq a}^N\sum_{n_a n_i} &\kf{a}{i} \bqty{\drdtheta{a}{i}}^2 \\
 &- \frac{\drv{a}{i}}{\rkikj{a}{i}} \frac{\pqty{\alpha-1}bn_a}{2I_1} \bqty{\frac{\pqty{\alpha-1}bn_a}{2} - \xkikj{a}{i}\sin\theta_a\sin\phi_a + \ykikj{a}{i}\sin\theta_a\cos\phi_a - \zkikj{a}{i}\cos\theta_a}
\end{aligned}
\end{align}
\begin{align}
\begin{aligned}
 \Melem{\tilde{\phi}}{a} = \sum_{i \neq a}^N\sum_{n_a n_i} &\kf{a}{i} \bqty{\drdphi{a}{i}}^2 \\
 &- \frac{\drv{a}{i}}{\rkikj{a}{i}} \frac{\pqty{\alpha-1}bn_a}{2I_1\sin^2\theta_a} \bqty{\frac{\pqty{\alpha-1}bn_a}{2}\sin^2\theta_a - \xkikj{a}{i}\sin\theta_a\sin\phi_a + \ykikj{a}{i}\sin\theta_a\cos\phi_a}
\end{aligned}
\end{align}
\begin{align}
 \Melem{x}[y]{a} = \sum_{i \neq a}^N\sum_{n_a n_i} \kf{a}{i} \frac{\xkikj{a}{i}\ykikj{a}{i}}{\rkikj{a}{i}^2}
\end{align}
\begin{align}
 \Melem{x}[z]{a} = \sum_{i \neq a}^N\sum_{n_a n_i} \kf{a}{i} \frac{\xkikj{a}{i}\zkikj{a}{i}}{\rkikj{a}{i}^2}
\end{align}
\begin{align}
 \Melem{y}[z]{a} = \sum_{i \neq a}^N\sum_{n_a n_i} \kf{a}{i} \frac{\ykikj{a}{i}\zkikj{a}{i}}{\rkikj{a}{i}^2}
\end{align}
\begin{align}
 \Melem{x}[\tilde{\theta}]{a} = \sum_{i \neq a}^N\sum_{n_a n_i} \kf{a}{i} \frac{\xkikj{a}{i}}{\rkikj{a}{i}} \bqty{\drdtheta{a}{i}} - \frac{\drv{a}{i}}{\rkikj{a}{i}} \frac{\pqty{\alpha-1}bn_a}{2\sqrt{I_1}} \cos\theta_a \sin\phi_a
\end{align}
\begin{align}
 \Melem{y}[\tilde{\theta}]{a} = \sum_{i \neq a}^N\sum_{n_a n_i} \kf{a}{i} \frac{\ykikj{a}{i}}{\rkikj{a}{i}} \bqty{\drdtheta{a}{i}} + \frac{\drv{a}{i}}{\rkikj{a}{i}} \frac{\pqty{\alpha-1}bn_a}{2\sqrt{I_1}} \cos\theta_a \cos\phi_a
\end{align}
\begin{align}
 \Melem{z}[\tilde{\theta}]{a} = \sum_{i \neq a}^N\sum_{n_a n_i} \kf{a}{i} \frac{\zkikj{a}{i}}{\rkikj{a}{i}} \bqty{\drdtheta{a}{i}} + \frac{\drv{a}{i}}{\rkikj{a}{i}} \frac{\pqty{\alpha-1}bn_a}{2\sqrt{I_1}} \sin\theta_a
\end{align}
\begin{align}
 \Melem{x}[\tilde{\phi}]{a} = \sum_{i \neq a}^N\sum_{n_a n_i} \kf{a}{i} \frac{\xkikj{a}{i}}{\rkikj{a}{i}} \bqty{\drdphi{a}{i}} - \frac{\drv{a}{i}}{\rkikj{a}{i}} \frac{\pqty{\alpha-1}bn_a}{2\sqrt{I_1}\sin\theta_a} \sin\theta_a \cos\phi_a
\end{align}
\begin{align}
 \Melem{y}[\tilde{\phi}]{a} = \sum_{i \neq a}^N\sum_{n_a n_i} \kf{a}{i} \frac{\ykikj{a}{i}}{\rkikj{a}{i}} \bqty{\drdphi{a}{i}} - \frac{\drv{a}{i}}{\rkikj{a}{i}} \frac{\pqty{\alpha-1}bn_a}{2\sqrt{I_1}\sin\theta_a} \sin\theta_a \sin\phi_a
\end{align}
\begin{align}
 \Melem{z}[\tilde{\phi}]{a} = \sum_{i \neq a}^N\sum_{n_a n_i} \kf{a}{i} \frac{\zkikj{a}{i}}{\rkikj{a}{i}} \bqty{\drdphi{a}{i}}
\end{align}
\begin{align}
\begin{aligned}
 \Melem{\tilde{\theta}}[\tilde{\phi}]{a} = \sum_{i \neq a}^N\sum_{n_a n_i} &\kf{a}{i} \bqty{\drdtheta{a}{i}} \\
 &\times \bqty{\drdphi{a}{i}} - \frac{\drv{a}{i}}{\rkikj{a}{i}} \frac{\pqty{\alpha-1}bn_a}{2I_1\sin\theta_a} \pqty{\xkikj{a}{i}\cos\theta_a\cos\phi_a + \ykikj{a}{i}\cos\theta_a\sin\phi_a}
\end{aligned}
\end{align}

\subsection{Off-diagonal blocks}
\begin{align}
 \Melem{x}{i}[j] = \sum_{n_i n_j} \kf{i}{j} \frac{-\xkikj{i}{j}^2}{\rkikj{i}{j}^2} + \frac{\drv{i}{j}}{\rkikj{i}{j}}
\end{align}
\begin{align}
 \Melem{y}{i}[j] = \sum_{n_i n_j} \kf{i}{j} \frac{-\ykikj{i}{j}^2}{\rkikj{i}{j}^2} + \frac{\drv{i}{j}}{\rkikj{i}{j}}
\end{align}
\begin{align}
 \Melem{z}{i}[j] = \sum_{n_i n_j} \kf{i}{j} \frac{-\zkikj{i}{j}^2}{\rkikj{i}{j}^2} + \frac{\drv{i}{j}}{\rkikj{i}{j}}
\end{align}
\begin{align}
\begin{aligned}
 \Melem{\tilde{\theta}}{i}[j] = \sum_{n_i n_j} &\kf{i}{j} \bqty{\drdtheta{i}{j}} \\
 &\times \bqty{\drdthetaj{i}{j}} \\
 &+ \frac{\drv{i}{j}}{\rkikj{i}{j}} \pqty{\frac{\alpha-1}{2}}^2 \frac{b^2n_in_j}{I_1} \pqty{\sin\theta_i\sin\theta_j + \cos\theta_i\cos\theta_j\cos\pqty{\phi_i - \phi_j}}
\end{aligned}
\end{align}
\begin{align}
\begin{aligned}
 \Melem{\tilde{\phi}}{i}[j] = \sum_{n_i n_j} &\kf{i}{j} \bqty{\drdphi{i}{j}} \\
 &\times \bqty{\drdphij{i}{j}} + \frac{\drv{i}{j}}{\rkikj{i}{j}} \pqty{\frac{\alpha-1}{2}}^2 \frac{b^2n_in_j}{I_1\sin\theta_i\sin\theta_j} \sin\theta_i\sin\theta_j\cos\pqty{\phi_i - \phi_j}
\end{aligned}
\end{align}
\begin{align}
 \Melem{x}[y]{i}[j] = \sum_{n_i n_j} \kf{i}{j} \frac{- \xkikj{i}{j} \ykikj{i}{j}}{\rkikj{i}{j}^2} = \Melem{y}[x]{i}[j]
\end{align}
\begin{align}
 \Melem{x}[z]{i}[j] = \sum_{n_i n_j} \kf{i}{j} \frac{- \xkikj{i}{j} \zkikj{i}{j}}{\rkikj{i}{j}^2} = \Melem{z}[x]{i}[j]
\end{align}
\begin{align}
 \Melem{y}[z]{i}[j] = \sum_{n_i n_j} \kf{i}{j} \frac{- \ykikj{i}{j} \zkikj{i}{j}}{\rkikj{i}{j}^2} = \Melem{z}[y]{i}[j]
\end{align}
\begin{align}
 \Melem{x}[\tilde{\theta}]{i}[j] = \sum_{n_i n_j} \kf{i}{j} \frac{\xkikj{i}{j}}{\rkikj{i}{j}} \bqty{\drdthetaj{i}{j}} + \frac{\drv{i}{j}}{\rkikj{i}{j}} \frac{\pqty{\alpha-1}bn_j}{2\sqrt{I_1}} \cos\theta_j\sin\phi_j
\end{align}
\begin{align}
 \Melem{\tilde{\theta}}[x]{i}[j] = \sum_{n_i n_j} \kf{i}{j} \frac{-\xkikj{i}{j}}{\rkikj{i}{j}} \bqty{\drdtheta{i}{j}} + \frac{\drv{i}{j}}{\rkikj{i}{j}} \frac{\pqty{\alpha-1}bn_i}{2\sqrt{I_1}} \cos\theta_i\sin\phi_i
\end{align}
\begin{align}
 \Melem{y}[\tilde{\theta}]{i}[j] = \sum_{n_i n_j} \kf{i}{j} \frac{\ykikj{i}{j}}{\rkikj{i}{j}} \bqty{\drdthetaj{i}{j}} - \frac{\drv{i}{j}}{\rkikj{i}{j}} \frac{\pqty{\alpha-1}bn_j}{2\sqrt{I_1}} \cos\theta_j\cos\phi_j
\end{align}
\begin{align}
 \Melem{\tilde{\theta}}[y]{i}[j] = \sum_{n_i n_j} \kf{i}{j} \frac{-\ykikj{i}{j}}{\rkikj{i}{j}} \bqty{\drdtheta{i}{j}} - \frac{\drv{i}{j}}{\rkikj{i}{j}} \frac{\pqty{\alpha-1}bn_i}{2\sqrt{I_1}} \cos\theta_i\cos\phi_i
\end{align}
\begin{align}
 \Melem{z}[\tilde{\theta}]{i}[j] = \sum_{n_i n_j} \kf{i}{j} \frac{\zkikj{i}{j}}{\rkikj{i}{j}} \bqty{\drdthetaj{i}{j}} - \frac{\drv{i}{j}}{\rkikj{i}{j}} \frac{\pqty{\alpha-1}bn_j}{2\sqrt{I_1}} \sin\theta_j
\end{align}
\begin{align}
 \Melem{\tilde{\theta}}[z]{i}[j] = \sum_{n_i n_j} \kf{i}{j} \frac{-\zkikj{i}{j}}{\rkikj{i}{j}} \bqty{\drdtheta{i}{j}} - \frac{\drv{i}{j}}{\rkikj{i}{j}} \frac{\pqty{\alpha-1}bn_i}{2\sqrt{I_1}} \sin\theta_i
\end{align}
\begin{align}
 \Melem{x}[\tilde{\phi}]{i}[j] = \sum_{n_i n_j} \kf{i}{j} \frac{\xkikj{i}{j}}{\rkikj{i}{j}} \bqty{\drdphij{i}{j}} + \frac{\drv{i}{j}}{\rkikj{i}{j}} \frac{\pqty{\alpha-1}bn_j}{2\sqrt{I_1}\sin\theta_j} \sin\theta_j \cos\phi_j
\end{align}
\begin{align}
 \Melem{\tilde{\phi}}[x]{i}[j] = \sum_{n_i n_j} \kf{i}{j} \frac{-\xkikj{i}{j}}{\rkikj{i}{j}} \bqty{\drdphi{i}{j}} + \frac{\drv{i}{j}}{\rkikj{i}{j}} \frac{\pqty{\alpha-1}bn_i}{2\sqrt{I_1}\sin\theta_i} \sin\theta_i \cos\phi_i
\end{align}
\begin{align}
 \Melem{y}[\tilde{\phi}]{i}[j] = \sum_{n_i n_j} \kf{i}{j} \frac{\ykikj{i}{j}}{\rkikj{i}{j}} \bqty{\drdphij{i}{j}} + \frac{\drv{i}{j}}{\rkikj{i}{j}} \frac{\pqty{\alpha-1}bn_j}{2\sqrt{I_1}\sin\theta_j} \sin\theta_j \sin\phi_j
\end{align}
\begin{align}
 \Melem{\tilde{\phi}}[y]{i}[j] = \sum_{n_i n_j} \kf{i}{j} \frac{-\ykikj{i}{j}}{\rkikj{i}{j}} \bqty{\drdphi{i}{j}} + \frac{\drv{i}{j}}{\rkikj{i}{j}} \frac{\pqty{\alpha-1}bn_i}{2\sqrt{I_1}\sin\theta_i} \sin\theta_i \sin\phi_i
\end{align}
\begin{align}
 \Melem{z}[\tilde{\phi}]{i}[j] = \sum_{n_i n_j} \kf{i}{j} \frac{\zkikj{i}{j}}{\rkikj{i}{j}} \bqty{\drdphij{i}{j}}
\end{align}
\begin{align}
 \Melem{\tilde{\phi}}[z]{i}[j] = \sum_{n_i n_j} \kf{i}{j} \frac{-\zkikj{i}{j}}{\rkikj{i}{j}} \bqty{\drdphi{i}{j}}
\end{align}
\begin{align}
\begin{aligned}
 \Melem{\tilde{\theta}}[\tilde{\phi}]{i}[j] = \sum_{n_i n_j} &\kf{i}{j} \bqty{\drdtheta{i}{j}} \\
 &\times \bqty{\drdphij{i}{j}} + \frac{\drv{i}{j}}{\rkikj{i}{j}} \pqty{\frac{\alpha-1}{2}}^2 \frac{b^2n_in_j}{I_1\sin\theta_j} \cos\theta_i\sin\theta_j\sin\pqty{\phi_i - \phi_j}
\end{aligned}
\end{align}
\begin{align}
\begin{aligned}
 \Melem{\tilde{\phi}}[\tilde{\theta}]{i}[j] = \sum_{n_i n_j} &\kf{i}{j} \bqty{\drdphi{i}{j}} \\
 &\times \bqty{\drdthetaj{i}{j}} - \frac{\drv{i}{j}}{\rkikj{i}{j}} \pqty{\frac{\alpha-1}{2}}^2 \frac{b^2n_in_j}{I_1\sin\theta_i} \sin\theta_i\cos\theta_j\sin\pqty{\phi_i - \phi_j}
\end{aligned}
\end{align}
\end{widetext}

\end{document}